\documentclass{article}

\usepackage[T1]{fontenc}
\usepackage[utf8]{inputenc}
\usepackage{graphicx}
\usepackage{amsmath}
\usepackage{dsfont}
\usepackage{txfonts}
\usepackage{multirow}
\usepackage{url}
\usepackage{color}
\usepackage{soul}
\usepackage{cancel}
\usepackage{pjournal}
\usepackage{jcappub}

\usepackage{braket}
\usepackage{color}
\usepackage{geometry}
\usepackage{xfrac}
\usepackage{bm}
\newcommand{\be}{\begin{equation}}
\newcommand{\ee}{\end{equation}}
\def\ba#1\ea{\begin{align}#1\end{align}}

\newcommand{\dd}{\mathrm{d}}

\newcommand{\hn}{\hat{n}}
\newcommand{\mr}{\mathrm}
\newcommand{\lbra}{\left\langle}
\newcommand{\rbra}{\right\rangle}
\newcommand{\Cov}{\mr{Cov}}

\newcommand{\Cl}{\ensuremath{C_\ell}}

\newcommand{\fsky}{\ensuremath{f_\mathrm{SKY}}}
\newcommand{\Euclid}{\textit{Euclid}}

\usepackage[normalem]{ulem}

\newcommand{\bbeta}{\bm{\beta}}
\newcommand{\bn}{\bm{n}}

\newcommand{\tz}{\tilde{z}}
\newcommand{\tbn}{\tilde{\bm{n}}}
\newcommand{\al}{\alpha}
\newcommand{\HH}{\mathcal{H}}
\newcommand{\ta}{\tilde{a}}

\newcommand{\tN}{\tilde{N}}
\newcommand{\lmax}{\ensuremath{\ell_\mathrm{max}}}
\newcommand{\Dkin}{\ensuremath{D_\mathrm{kin}}}
\newcommand{\Fkin}{\ensuremath{F_\mathrm{kin}}}

\newcommand{\hbeta}{\hat{\beta}}
\newcommand{\hF}{\hat{F}_\mr{kin}}
\newcommand{\dg}{\delta_\mr{g}}

\newcommand{\threej}[6]{\begin{pmatrix}
#1 & #2 & #3 \\ #4 & #5 & #6
\end{pmatrix}}
\newcommand{\threejzero}[3]{\begin{pmatrix}
#1 & #2 & #3 \\ 0 & 0 & 0
\end{pmatrix}}

\definecolor{dgreen}{rgb}{0,0.6,0}

\begin{document}

\title{Fast and spurious: a robust determination of our peculiar velocity with future galaxy surveys}

\author[a,b]{Fabien Lacasa}
\author[a]{Camille Bonvin}
\author[c,d]{Charles Dalang}
\author[a]{Ruth Durrer}

\affiliation[a]{D\'{e}partement de Physique Th\'{e}orique and Center for Astroparticle Physics, Universit\'{e} de Gen\`{e}ve, 24 quai Ernest Ansermet, CH-1211 Geneva, Switzerland}
\affiliation[b]{Université Paris-Saclay, CNRS, Institut d’Astrophysique Spatiale, 91405, Orsay, France}
\affiliation[c]{Queen Mary University of London, Mile End Road, E1 4NS, London, United Kingdom}
\affiliation[d]{Institute of Cosmology and Gravitation, University of Portsmouth, Burnaby Road, Portsmouth PO1 3FX, United Kingdom}

\emailAdd{fabien.lacasa@unige.ch}

\date{}

\abstract{
To date, the most precise measurement of the observer's peculiar velocity comes from the dipole in the Cosmic Microwave Background (CMB).
This velocity also generates a dipole in the source number counts, whose amplitude is governed not only by the observer velocity, but also by specific properties of the sources, that are difficult to determine precisely. Quantitative studies of the source number counts currently give dipoles which are reasonably well aligned with the CMB dipole, but with a significantly larger amplitude than that of the CMB dipole. In this work, we explore an alternative way of measuring the observer velocity from the source number counts, using correlations between neighboring spherical harmonic coefficients, induced by the velocity. We show that these correlations contain both a term sensitive to the source properties and another one directly given by the observer velocity. We explore the potential of a \Euclid-like survey to directly measure this second contribution, independently of the characteristics of the population of sources. We find that the method can reach a precision of 4\%, corresponding to a detection significance of $24\sigma$, on the observer velocity. This will settle with precision the present "dipole tension". 
}

        
\maketitle

%%%%%%%%%%%%%%%%%%%%%%%%%%%%%%%%%%%%%%%%%%%%%%%%%%%%%%%%%%%%%%%%%%%
\section{Introduction}\label{Sect:intro}

The dipole of the Cosmic Microwave Background (CMB) has been measured with outstanding precision. Since its amplitude is about 100 times larger than all higher multipoles, it is usually interpreted as due to our peculiar motion with respect to the cosmic rest frame. Such a motion does indeed generate a dipole $\bm{D}_\mr{kin}^\mr{CMB} =\bm{v}_o/c$ at linear order in the velocity $v_o$ \cite{Peebles:1968}.
Attributing the observed dipole entirely to our motion with respect to the cosmic rest frame, one infers a velocity amplitude and direction of~\cite{Kogut:1993ag,Planck:2018nkj}
\be
|\bm{v}_0|= (369.82 \pm 0.11) \ \mr{km/s}\,, \qquad %(369\pm 0.9)\mr{km/s} \,, \qquad  
(l,b) = (264.021^\circ \pm 0.011^\circ, 48.253^\circ \pm 0.005^\circ)\, , %(263.99\pm 0.14, 48.26\pm 0.03)
\label{Eq:d-Planck}
\ee
where $(l,b)$ denote the longitude and latitude of the velocity direction in galactic coordinates. Assuming that the kinematic dipole is contaminated by an intrinsic dipole of a similar amplitude as the higher multipoles induces an error of about 1\% in the above determination of $\bm{v}_0$.

In addition to the dipole in the CMB temperature, our motion is expected to generate a dipole in the distribution of cosmological sources around us. In recent years, several groups 
were able to detect this dipole in cosmic sources, observing radio galaxies~\cite{Bengaly:2017slg,Tiwari:2016,Siewert:2021,Wagenveld:2023kvi,Tiwari:2018hrs,Colin:2019opb}
and, more recently, quasars~\cite{Secrest:2020has,Secrest:2022uvx,Dam:2022wwh}. Neglecting the redshift evolution of the sources, the kinematic dipole of radio galaxies and quasars, averaged over all redshifts (which is denoted here only, with a bar), can be written as \cite{Ellis1984}
\be
\bar{\bm{D}}_\mr{kin} = [2+x(1+\alpha)] \ \bm{v}_o/c \,, 
\label{Eq:EB}
\ee
where $x= -\dd\ln N (S>S_*)/\dd\ln S_{*}$ is the slope of the cumulative number of sources at the limiting flux of the survey $S_*$, and $\alpha =-\dd\ln S/\dd\ln f\mid_{f_o}$ is the spectral index of the sources evaluated at the observed frequency $f_o$. Eq.~\eqref{Eq:EB} is the so-called \textit{Ellis \& Baldwin} formula~\cite{Ellis1984}.
Surprisingly, measurements of the kinematic dipole of sources, while not very well agreeing with each other (see e.g.\,\cite{Cheng2023}), find a dipole with a direction roughly compatible with that of the CMB, but with an amplitude significantly larger than expected \cite{Bengaly:2017slg,Tiwari:2016,Siewert:2021,Wagenveld:2023kvi,Tiwari:2018hrs,Colin:2019opb}. Recently, the authors of~\cite{Secrest:2020has,Secrest:2022uvx}, analysing a large sample of quasars, have obtained a result which is $5\sigma$ away from the value predicted from the CMB dipole. In particular, using Eq.~\eqref{Eq:EB}, they inferred an observer's velocity which is twice as large as in Eq.\,\eqref{Eq:d-Planck}. These results hold when one accounts for the presence of the Galactic plane mask, which leads to a leakage of power from the dipole into higher multipoles, further increasing the tension with the  CMB~\cite{Dam:2022wwh}. The authors consider that their finding is a true challenge for the cosmological standard model.
This conclusion, as well as the significance of the tension with respect to the CMB, relies however entirely on the ability to predict the value of the prefactor $[2+x(1+\alpha)]$ in Eq.\,\eqref{Eq:EB}, since
$\bm{v}_o$ cannot be measured directly, but only inferred from Eq.\,\eqref{Eq:EB}. In particular, it has been shown that the redshift dependence of $\al$ and $x$ might significantly affect the result if these quantities are correlated~\cite{Dalang2021,Guandalin2022}. Of course, the importance of this effect depends on the strength of the correlation between the parameters~\cite{Dam:2022wwh}, which is difficult to assess.
In the near future, measurements of the near-IR Cosmic Infrared Background (CIB) with \Euclid{} promise to constrain the direction of the dipole to sub-degree precision \cite{Kashlinsky2024} ; the interpretation of the amplitude will however be weak to the same caveats as for the radio galaxies and quasars.

An independent avenue to shed light on this tension is to use the fact that our peculiar velocity introduces a preferred direction of observation in the sky and consequently breaks statistical isotropy. This effect leads to a mode coupling in the CMB harmonic coefficients leading to non-vanishing correlations between $a_{\ell, m}$ and $a_{\ell\pm 1,m }$. These correlations have been detected in the CMB with high significance~\cite{Planck:2013kqc,Saha:2021bay,Ferreira:2020aqa}, and the inferred velocity $v_o = (384 \pm 78 (\hbox{stat.}) \pm 115 (\hbox{syst.}))$\,km$/$s is compatible with Eq.\,\eqref{Eq:d-Planck} although with much larger 
error bars. 

The same effect can also be searched for in galaxy surveys, and Ref.\,\cite{Dalang2022} showed that, compared to the CMB, the higher angular resolution of galaxy surveys yields better prospects for precise measurements of the observer velocity with this method. 
In~\cite{Dalang2022}, it was found that the correlations between $a_{\ell, m}$ and $a_{\ell\pm 1,m }$ in galaxy surveys are governed by the same combination $\Dkin$ as the dipole. In this article, we revisit this computation. We show that three terms were missing in the calculation of~\cite{Dalang2022}. The first two terms have the same multipole dependence as in~\cite{Dalang2022}, they thus renormalise $\Dkin$ into a new quantity that we call $\Fkin$. The last term is very interesting, since it is directly proportional to the observer's velocity $\bm{v}_o$, with no dependence on the sources properties. Moreover it does not scale in the same way with multipole $\ell$ as the contribution proportional to $\Fkin$.
This allows us to efficiently break the degeneracy between $\bm{v}_o$ and the source population parameters. As a consequence, the correlations between $a_{\ell, m}$ and $a_{\ell\pm 1,m }$ (that we call hereafter $(\ell, \ell\pm 1))$ correlations) provide a robust way of measuring $\bm{v}_o$ without needing a prior knowledge of the properties of the sources. This is particularly important, because precisely measuring the source parameters is not trivial. Our calculation accounts for the evolution of the $(\ell, \ell\pm 1)$ correlations with redshift and is therefore applicable to the case of a tomographic survey, with various redshift bins.

We forecast the detectability of $v_o$ using the $(\ell, \ell\pm 1))$ correlations, for a \Euclid-like photometric galaxy survey. Combining the 10 redshift bins of \Euclid{}, we find that the observer velocity can be detected at nearly 3$\sigma$ if its amplitude is as predicted from the CMB dipole. The significance of the detection is limited by the degeneracy between the contribution proportional to $v_o$ and the contribution proportional to $\Fkin$. We show that this degeneracy can be strongly reduced if we assume some prior knowledge on $\Fkin$. Interestingly, with a prior of only 30\% on the source parameters that enter into $\Fkin$, we reach a detection significance of $9\sigma$, corresponding to a precision of 11\% on the measurement of $v_o$. If we tighten the priors to 10\%, we reach a significance of $24\sigma$, corresponding to a precision of 4\%. Importantly, by building an optimal estimator for $v_o$,  we show that our method does not rely on the modelling of the unboosted angular power spectrum of galaxies, $C_{\ell}$. The estimator is instead directly built from the measured $C_{\ell}$. For this reason, the determination of $v_o$ does not depend on our ability to model non-linear physics.

The present article is structured as follows: In Section~\ref{Sect:boost-observables}, we derive the impact of the observer velocity on the perturbed galaxy number counts, accounting for all terms of the form $v_o\times \delta_g$, where $\delta_g$ denotes the galaxy density perturbation. In Section~\ref{Sect:stat-impact}, we derive the correlations between the $\ell$ and $\ell\pm 1$ spherical harmonic coefficients. In particular, we show the appearance of an extra term directly proportional to $v_o$ in the $(\ell,\ell+1)$ correlators which was missed in~\cite{Dalang2022}. In Section~\ref{Sect:estimators}, we derive the maximum likelihood estimator for $\Fkin{}(z)$ and $v_o$ using the $(\ell,\ell+1)$  correlators, and we study the degeneracy between $\Fkin{}(z)$ and $v_o$. We then show how this degeneracy can be lifted by the study of correlations of different redshift bins or of a multi-tracer analysis.
Finally, in Section~\ref{Sect:forecast}, we forecast the prospects for a \Euclid-like galaxy survey and potential strategies to further improve the measurement precision. Several more technical aspects are relegated to the appendices. 

We work in units in which the speed of light $c=1$ such that $\beta\equiv v_o/c=v_o$. We consider a spatially flat perturbed Friedmann-Lemaître Universe with background metric $\dd s^2 = a^2(\tau)[-\dd \tau^2 +\delta_{ij} \dd x^i \dd x^j]$. An overdot denotes a derivative with respect to conformal time $\tau$, which relates to cosmic time $t$ via the scale factor $a(\tau)$: $a(\tau) \dd \tau = \dd t $. The conformal Hubble rate is defined as $\mathcal{H} = \dot{a}/a = Ha$, where $H$ is the physical Hubble rate. We denote the Kronecker delta by $\delta^\mr{K}_{ij}$.

%%%%%%%%%%%%%%%%%%%%%%%%%%%%%%%%%%%%%%%%%%%%%%%%%%%%%%%%%%%%%%%%%%%

\section{Effect of the observer velocity on the perturbed galaxy number counts}\label{Sect:boost-observables}

In this section, we derive the relation between the boosted and unboosted galaxy number counts. We adapt the calculations presented in \cite{Maartens:2017qoa,Dalang2021} and work out the details of the calculation sketched in~\cite{Dalang2022}, which misses an important term. We then compute the harmonic coefficients of the observed boosted number counts.

%%%%%%%%%%%%%%%%%%%%
\subsection{Boosted galaxy number counts}
\label{Sect:boost}

Let us denote by $\tilde N(\tilde z, \tilde\bn)$ the unboosted number density of galaxies per redshift bin $ \dd \tilde z$ and per solid angle $\dd \tilde\Omega$ in direction $\tilde\bn$ and at redshift $\tilde z$.
Correspondingly, we denote the boosted, observed quantity $N(z,\bn)$. We choose coordinates such that the boost $\bm{\beta} \equiv \bm{v}_o/c$ is aligned with the $+\bm{e}_z$ axis. As shown in~\cite{Maartens:2017qoa,Nadolny:2021hti,Dalang2021}, the dipole in the distribution of galaxies is due to the impact of the boost on the monopole, the isotropic part of the distribution of galaxies $\tilde{\bar{N}}(\tilde{\bar{z}})$. The correlations between the harmonic coefficients, $a_{\ell, m}$ and $a_{\ell\pm 1,m }$, on the other hand are due to the effect of the boost on the perturbed anisotropic distribution of galaxies $\tilde N(\tilde z, \tilde\bn)$. In our derivation, we keep therefore terms that are linear in $v_o$ and that multiply the perturbed galaxy distribution. These terms are formally beyond linear order. As shown in~\cite{Bonvin2011}, various terms contribute to the perturbed galaxy distribution at linear order. Calculating the impact of the boost on all those terms is beyond the scope of this article. Instead we focus on the dominant contribution to the perturbed galaxy distribution, namely the density fluctuation $\delta_{\rm g}=b\cdot\delta_\mr{m}$, where $b$ is the galaxy bias, and $\delta_{\rm m}$ is the matter density contrast. We neglect redshift-space distortions, since we will use photometric galaxy surveys with relatively wide redshift bins, where redshift-space distortions are strongly suppressed. We also neglect the effect of lensing magnification. This term is completely negligible at low redshift, but it starts to become relevant at high redshift~\cite{Euclid:2021rez}. Neglecting it may therefore slightly affect the constraints in the highest redshift bins of \textit{Euclid}, but we do not expect this to change the overall results. Finally, we also neglect the terms that are quadratic in the observer velocity, and which are suppressed with respect to terms of the form $v_o\cdot \delta_\mr{m}$, since $v\sim \frac{\mathcal{H}}{k} \delta_\mr{m} \ll \delta_\mr{m}$ for sub-horizon scales.

We now proceed to the calculation of the boosted number counts in terms of the unboosted number counts. The boosted redshift $z$, direction $\bn = (\theta,\phi)$ and solid angle $\dd \Omega$ are related to their unboosted counterparts, denoted with tilde, by the following equations, see e.g.~\cite{Dalang2021}
\begin{align}
1 + \tilde z & = (1+ z)(1+\bbeta\cdot\bn)\,, \label{Eq:redshift}
\\
\tilde\bn & = (1+\bn\cdot\bbeta)\bn-\bbeta\,, \label{Eq:n}
\\
\sin\tilde\theta & = (1+\bn\cdot\bbeta)\sin\theta\,, \label{Eq:sin}
\\
\tilde\theta & = \theta + \beta\sin\theta + \mathcal{O}\left(\beta^2\right)\,,\label{Eq:sin_theta}
\\
\dd \tilde\Omega & = \dd \Omega \times \left( 1 + 2 \, \bn\cdot\bbeta + \mathcal{O}\left(\beta^2\right) \right)  \,. \label{Eq:solid_angle}
\end{align}
The number of galaxies is conserved under the effect of the boost: all boosted galaxies observed in a solid angle $\dd \Omega$ and redshift bin $\dd z$ with flux above $S_*$, would be observed in solid angle $\dd \tilde{\Omega}$ and redshift bin $\dd \tilde{z}$ if the observer were not moving. 
Moreover, these galaxies all have an intrinsic luminosity larger than $L_*(z,\bn)=4\pi d_L^2(z,\bn)S_*$, where $d_L$ denotes the luminosity distance. As a consequence, we can write
\begin{align}
N(z,\bm{n},S_*) \ \dd z \; \dd \Omega = N(z,\bm{n},L_*(z,\bn)) \ \dd z \; \dd \Omega=\tilde N (\tilde z , \bm{\tilde n}, L_*(z,\bn)) \ \dd \tilde z \; \dd \tilde \Omega \,,\label{Eq:Nconservation2}
\end{align}
where we shortened notations for simplicity with $N(z,\bm{n},S_*) \equiv N(z,\bm{n},S>S_*)$, and similarly for $N(z,\bm{n},L_*(z,\bn))$.
Dividing both sides of the equation by $\dd\Omega \dd z$ and Taylor expanding $\tilde{N}$ around the unboosted luminosity threshold $\tilde{L}_*$ (i.e.\ the luminosity threshold that would correspond to $S_*$ if the observer was not moving), we obtain
\begin{align}
N(z,\bm{n}, S_*) = \left[\tilde{N}(\tilde{z},\tilde{\bn}, \tilde{L}_*(\tilde{z},\tilde{\bn}))+\left(L_*(z,\bn)-\tilde{L}_*(\tilde{z},\tilde{\bn})\right) \frac{\partial}{\partial \tilde{L}_*}\tilde{N}(\tilde{z},\tilde{\bm{n}}, \tilde{L}_*)\right]\frac{\dd \tilde{\Omega}}{\dd \Omega}\frac{\dd \tilde{z}}{\dd z} \,. \label{Eq:Nobserved}
\end{align}
The luminosity derivative reads to zeroth order in $\beta$
\begin{align}
\frac{\partial}{\partial \tilde{L}_*}\tilde{N}(\tilde{z},\tilde{\bm{n}},\tilde{L}_*) = -\frac{x(\tilde{z})}{\tilde{L}_*}\tilde{N}(\tilde{z},\tilde{\bm{n}}, \tilde{L}_*)
+ \frac{\tilde{\bar{N}}(\tilde{z},\tilde{L}_*)}{\tilde{L}_*}\frac{\partial}{\partial\ln \tilde{L}_*}\left(\frac{\delta\tilde{N}(\tilde{z},\tilde{\bm{n}},\tilde{L}_*)}{\tilde{\bar{N}}(\tilde{z},\tilde{L}_*)}\right)\,, \label{Eq:luminosity_derivative}
\end{align}
where $\tilde{\bar{N}}$ denotes the average number of galaxies at redshift $\tilde{z}$, $\delta\tilde{N}$ is the departure from the average, and the average magnification bias is defined as $x(\tilde{z})\equiv-\partial\ln \tilde{\bar{N}}/\partial\ln \tilde{L}_*$. \\
As the flux cut $S_*$ is a property of the survey, independent of the boost, it corresponds to different luminosity cuts for the boosted and unboosted observer,
\begin{align}
 S_*=\frac{\tilde{L}_*(\tz,\tilde{\bn})}{4\pi \tilde{d}^2_L(\tilde{z},\tilde{\bn})} =\frac{L_*(z,\bn)}{4\pi d_L^2(z,\bn)}\,,
 \label{Eq:flux_distance}
\end{align}
from which we obtain to first order in perturbations
\begin{align}
\frac{L_*(z,\bm{n})-\tilde{L}_*(\tz,\tbn)}{\tilde{L}_*(\tz,\tbn)}= \frac{2 \left(d_L(z,\bm{n})-\tilde{d}_L(\tz,\tbn)\right)}{\tilde{d}_L(\tz,\tbn)}\,.
\end{align}
The impact of the observer's velocity on the luminosity distance at fixed observed redshift is given by~\cite{Bonvin:2005ps}
\begin{align}
\frac{\left(d_L(z,\bm{n})-\tilde{d}_L(\tz,\tbn)\right)}{\tilde{d}_L(\tz,\tbn)} =  \frac{\bm{n}\cdot \bm{\beta}}{r \mathcal{H}}\,, \label{Eq:dL_vel}
\end{align}
where $r$ is the comoving distance at redshift $z$ and $\HH$ is the conformal Hubble parameter.
Combining Eqs.\,\eqref{Eq:redshift}, \eqref{Eq:solid_angle}, \eqref{Eq:Nobserved}, \eqref{Eq:luminosity_derivative} and \eqref{Eq:dL_vel}, we obtain to first order in $\beta$
\begin{align}
N(z,\bm{n}, S_* ) = \left (1+3 \bm{n}\cdot \bm{\beta} \right ) \tilde N\left(\tilde z,\bm{\tilde n},\tilde{L}_*(\tz,\tbn)\right) +  \frac{2\bm{n}\cdot \bm{\beta}}{r\mathcal{H}} \left[-x(\tilde{z})\tilde N \left(\tilde{z},\tilde{\bm{n}},\tilde{L}_*(\tz,\tbn)\right)+\tilde{\bar{N}}(\tilde{z},\tilde{L}_*)\frac{\partial}{\partial\ln \tilde{L}_*}\left(\frac{\delta\tilde{N}(\tilde{z},\tilde{\bn},\tilde{L}_*)}{\tilde{\bar{N}}(\tilde{z},\tilde{L}_*)}\right)\right]\,.
\label{Eq:N_interm}
\end{align}

The quantity $\tilde{N}\left(\tilde z,\bm{\tilde n},\tilde{L}_*(\tz,\tbn)\right)$, which appears on the right-hand side of Eq.\,\eqref{Eq:N_interm}, is the number of galaxies detected by an unboosted observer, at rest with respect to the CMB frame. From Eq.\,\eqref{Eq:flux_distance}, we see that all galaxies $\tilde{N}$ that have intrinsic luminosity larger than $\tilde{L}_*$ in the unboosted CMB frame, would be observed with a flux $S>S_*$ in that frame. Hence we can rewrite $\tilde{N}\left(\tilde z,\bm{\tilde n},\tilde{L}_*(\tz,\tbn)\right)=\tilde{N}\left(\tilde z,\bm{\tilde n},S_*\right)$. This quantity depends on the unboosted redshift $\tilde{z}$ and the unboosted direction $\tilde{\bn}$, that are not observable: one only measures the boosted redshift $z$ and angular direction $\bm{n}$. Hence to predict the value of $N(z,\bm{n}, S_* )$ in terms of observable quantities, it is necessary to further Taylor-expand $\tilde{N}\left(\tilde z,\bm{\tilde n},S_*\right)$ around $z$ and $\bn$. To first order in $\beta$, we have
\be
\tilde N(\tilde z, \tilde\bn, S_*) = \tilde N(z, \bn, S_*) + (\tilde \bn -\bn)\frac{\partial \tilde N(z, \bn,S_*)}{\partial \bn} + (\tilde z -z)\frac{\partial \tilde N(z, \bn,S_*)}{\partial z} \,. \label{Eq:Taylor_Expansion_Ntilde}
\ee
Using Eq.~\eqref{Eq:redshift}, the third term can be written as
\be
(\tilde z -z)\frac{\partial  \tilde N(z, \bn,S_*)}{\partial z} = \bn \cdot \bbeta \frac{\partial \tilde N (z,\bn,S_*)}{\partial \ln(1+z)}\,. \label{Eq:remappingz}
\ee
We first write
\begin{align}
\tilde N (z,\bn,S_*)=\tilde{\bar{N}} (z,S_*) \ \left(1+\frac{\delta\tilde{N}(z,\bn,S_*)}{\tilde{\bar{N}}(z,S_*)} \right) \, ,   \label{eq:Nsplit}
\end{align}
and then we express the background number density (per unit solid angle and per redshift bin) in terms of the background comoving number density $n_\mr{c}(z,S_*)$. Using that $\dd \Omega \, \dd z = (1+z) \, \left(\mathcal{H}/r^2\right) \, \dd V_\mr{c}$, we obtain
\begin{align}
\tilde{\bar{N}} (z, S_*)= \frac{r^2}{\mathcal{H}(1+z)}  \ n_\mr{c}(z,S_*)\, .\label{eq:ncom}
\end{align}
Evaluating the logarithmic redshift derivative of Eq.~\eqref{eq:Nsplit}, using Eq.~\eqref{eq:ncom} we obtain
\be
\frac{\partial \tilde N(z,\bn,S_*)}{\partial \ln(1+z)} =\left[\frac{2}{r(z)\HH(z)} -1 +\frac{\dot \HH}{\HH^2} -f_\mr{evol}(z)\right] \tilde{N}(z,\bn,S_*) +\frac{\partial}{\partial\ln (1+z)}\left(\frac{\delta\tilde{N}(z,\bn,S_*)}{\tilde{\bar{N}}(z,S_*)}\right)\tilde{\bar{N}}(z,S_*)\,. \label{eq:variationN}
\ee
Here, the overdot denotes the derivative with respect to conformal time and we have introduced the so-called evolution bias
\be
f_\mr{evol}(z) \equiv - \frac{\partial \ln n_\mr{c}(z,S_*)}{\partial \ln(1+z)} \,.
\ee
Inserting this into Eq.\,\eqref{Eq:Taylor_Expansion_Ntilde} and using Eq.\,\eqref{Eq:sin_theta} to rewrite $(\tilde\bn-\bn )$, we obtain 
\begin{align}
N(z,\bn,S_*) =& \left [1+\cos\theta \, D_\mr{kin}(z)\right ] \tilde{\bar N}(z, S_* ) + \left [1+\cos\theta \, \Fkin(z,\bn)+\beta\sin\theta\,\partial_\theta\right ]\tilde{\bar N}(z, S_* ) \ \delta_\mr{g}(z,\bn,S_*)\, ,\label{Eq:Nfinal}
\end{align}
where
\begin{align}
\delta_\mr{g}(z,\bn,S_*)= \frac{\delta\tilde{N}(z,\bn,S_*)}{\tilde{\bar{N}}(z,S_*)}\, ,  
\end{align}
and
\begin{align}
D_\mr{kin}(z) &= \left[ 2 + \frac{2(1-x(z))}{r\HH}+\frac{\dot \HH}{\HH^2} -f_\mr{evol}(z)\right] \beta \,. \label{Eq:Dkinz}\\
\Fkin(z,\bn) &= D_\mr{kin}(z)+\left[\frac{\partial \ln\delta_\mr{g}}{\partial\ln (1+z)}+ \frac{2}{r\HH}\frac{\partial \ln\delta_\mr{g}}{\partial\ln S_*}\right]\beta\, . \label{Eq:Fkin_gen}
\end{align}
At linear order, the galaxy perturbations, $\delta_{\rm g}$, can be written as the product of a scale-independent bias, which depends on redshift and flux threshold, and the matter density fluctuations: $\delta_{\rm g}=b(z,S_*) \ \delta_\mr{m}(z,\bn)$. In this regime $\Fkin$ depends only on redshift and can be written as
\begin{align}
\Fkin(z)= D_\mr{kin}(z)+\left[- f(z)+\frac{\partial \ln b}{\partial\ln (1+z)}+ \frac{2}{r\HH}\frac{\partial \ln b}{\partial\ln S_*}\right]\beta\, ,  \label{Eq:Fkin}
\end{align}
where $f\equiv \dd\ln \delta_\mr{m}(a,\bn)/\dd\ln a$ is the growth rate of structure, which is scale-independent in the linear regime.

In the highly non-linear regime, the bias and growth rate would get scale-dependent corrections. We do not consider these in the main text, although we explain in Appendix~\ref{App:revising-assumptions-NLbg} how one could consider them and show that they should not impact the conclusions of the article.

The first contribution in Eq.~\eqref{Eq:Nfinal}, proportional to the background number of galaxies $\tilde{\bar N}(z, S_* )$ generates only a dipolar modulation in the boosted  number of galaxies. As already shown in~\cite{Maartens:2017qoa,Nadolny:2021hti,Dalang2021}, the amplitude of the dipole is governed by the combination $\Dkin(z)$ given in~\eqref{Eq:Dkinz}, which depends on the magnification bias and evolution bias of the population of galaxies. The second contribution in Eq.~\eqref{Eq:Nfinal}, proportional to the galaxy fluctuations, $\delta_\mr{g}(z,\bn)$, generates correlations between $a_{\ell, m}$ and $a_{\ell\pm1, m}$. We see that this contribution contains two terms: one proportional to $\cos\theta$, and a second term which depends on the variation of the galaxy fluctuations with $\theta$. This second term is a remapping contribution: it follows directly from the remapping of $\tN$ from the unboosted direction $\tbn$ to the boosted observed direction $\bn$. 

Comparing the contribution proportional to $\tilde{\bar N}(z, S_* )$ with the contribution proportional to $\dg(z,\bn)$ in Eq.~\eqref{Eq:Nfinal}, we see that the second one is strongly suppressed with respect to the first one, since $\dg\ll 1$. Hence, the dipolar modulation is completely dominated by the first contribution, proportional to $\Dkin$. The correlations between $a_{\ell, m}$ and $a_{\ell\pm1, m}$, on the other hand, are only due to the second contribution. What is very interesting, is that this second contribution has two terms, $\Fkin$ and the remapping, which do not have the same angular dependence in $\theta$. As we will see in Sec.~\ref{Sect:Spherical_harmonic_coefficient}, this will allow us to separate these two contributions, and measure separately $\Fkin(z)$ and $\beta$. This provides a strong advantage with respect to measurements of the dipole, since $\beta$ can be measured without requiring a model for the characteristics of the sources. Finally, let us note that $\Dkin$ and $\Fkin$ are not exactly the same. $\Fkin$ contains additional contributions, which depend on the evolution of the galaxy fluctuations with redshift and with flux cut.

%%%%%%%%%%%%%%%%%%%%
\subsection{Spherical harmonic coefficients}\label{Sect:Spherical_harmonic_coefficient}

In order to find the effect of a boost on the statistical distribution of galaxies, we decompose the number density of galaxies in spherical harmonics 
\begin{align}
a_{\ell,m}(z) = \int_{\mathbb{S}_2} \dd^2 \bn\, N(z,\bn, S_*) Y^{*}_{\ell,m}(\bn)\,.
\end{align}
In the following, we leave the dependence on the flux threshold and redshift implicit.
We also decompose the unboosted number counts in spherical harmonics, with coefficients $\tilde{a}_{\ell, m}$ indicated with a tilde. The statistics of $\ta_{\ell, m}$ can be predicted from a theoretical model of galaxy clustering. Our goal is to relate the observed coefficients $a_{\ell, m}$ to the underlying $\ta_{\ell, m}$. We have 
\ba
a_{\ell,m} =& \int_{\mathbb{S}_2} \dd^2\bn \ N(z,\bn) \ Y^*_{\ell,m}(\bn) \label{Eq:a_lm} \\
=& \Dkin(z)\,\tilde{\bar{N}}(z)\int_{\mathbb{S}_2} \dd^2\bn \cos\theta\, Y^*_{\ell, m}(\bn)+ \tilde{\bar{N}}(z)\int_{\mathbb{S}_2} \dd^2\bn \ \left[1+\cos\theta \, F_\mr{kin}(z)  + \beta\sin\theta \, \partial_{\theta}\right] \dg(z,\bn) \ Y^*_{\ell, m}(\bn) \nonumber \\
=& \ta_{\ell,m} + \sqrt{\frac{4\pi}{3}} \Dkin(z)\, \tilde{\bar{N}}(z)\, \delta^\mr{K}_{\ell,1}\,\delta^\mr{K}_{m,0} + \Fkin(z)\, \tilde{\bar{N}}(z)\int_{\mathbb{S}_2} \dd^2 \bn \ \dg(z,\bn) \ \cos\theta \ Y^*_{\ell,m}(\bn)\\
& - \beta\, \tilde{\bar{N}}(z)\int_{\mathbb{S}_2}  \dd^2 \bn \, \dg(z,\bn) \left( \sin\theta \, \partial_{\theta} Y^*_{\ell,m}(\bn) +2 \cos\theta \, Y^*_{\ell, m}(\bn)\right)
\nonumber
\ea
where we used that $\cos\theta = \sqrt{\frac{4\pi}{3}} Y_{1,0}(\bn)$ and we have performed an integration by part in the last line. Therefore, we write
\ba
a_{\ell,m} &= \ta_{\ell,m} + \delta a_{\ell,m}^\mr{Dip}+ \delta a_{\ell,m}^\mr{F} - \delta  a_{\ell,m}^\mr{Remap}\,, \label{Eq:linkalm}
\ea
with
\ba
\delta a_{\ell,m}^\mr{Dip}&=\sqrt{\frac{4\pi}{3}} \Dkin(z)\, \tilde{\bar{N}}(z)\, \delta^\mr{K}_{\ell 1}\,\delta^\mr{K}_{m 0}\, ,\label{Eq:alm_dip}\\
\delta a_{\ell,m}^\mr{F} &= \Fkin(z)\, \tilde{\bar{N}}(z) \int_{\mathbb{S}_2} \dd^2 \bn \ \dg(z,\bn) \ \cos\theta \ Y^*_{\ell, m}(\bn) \,, \label{Eq:alm_F}\\
\delta  a_{\ell,m}^\mr{Remap} &= \beta\, \tilde{\bar{N}}(z) \int_{\mathbb{S}_2} \dd^2 \bn \, \dg(z,\bn) \left( \sin\theta \, \partial_{\theta} Y^*_{\ell,m}(\bn) +2 \cos\theta \, Y^*_{\ell,m}(\bn)\right)\,. \label{Eq:alm_remap}
\ea
The minus sign in Eq.~\eqref{Eq:linkalm} has been introduced for convenience. The superscript ``Dip'' denotes the dipole contribution, which affects only the multipole $\ell=1, m=0$; the ``F'' superscript refers to the term proportional to $\Fkin$, while the ``Remap'' refers to remapping contribution, directly proportional to $\beta$. Using that the monopole $\int \dd^2\bn \, \tN(z,\bn) \, Y_{0,0}(\bn) =\sqrt{4\pi}\, \tilde{\bar{N}}(z)=\ta^\mr{BG}_{0,0}$, the dipole term in Eq.~\eqref{Eq:alm_dip} becomes
\begin{align}
\delta a_{\ell,m}^\mr{Dip}=\frac{\Dkin(z)}{\sqrt{3}}\ta^\mr{BG}_{0,0} \, \delta^\mr{K}_{\ell 1}\,\delta^\mr{K}_{m0}\,,
\end{align}
where we have introduced the upper script ``BG'' to emphasise that this term is the only one that depends on the background number of galaxies $\tilde{\bar{N}}(z)$, averaged over direction. All the other terms depends on the galaxy fluctuations $\dg$.
The F term in Eq.~\eqref{Eq:alm_F} can be rewritten, using calculations from Appendix~\ref{Sect:dipole-X-Ylm}
\ba
\delta a_{\ell,m}^\mr{F} &= \Fkin(z)\, \tilde{\bar{N}}(z) \int_{\mathbb{S}_2} \dd^2 \bn \, \dg(z,\bn)  \left[ A_{\ell,m} \, Y^*_{\ell-1,m}(\bn) + A_{\ell+1,m} \, Y^*_{\ell+1,m}(\bn) \right] \\
&= \Fkin(z) \ \left[ A_{\ell,m} \, \ta_{\ell-1,m} + A_{\ell+1,m} \, \ta _{\ell+1,m} \right]\,, \label{Eq:alm_F_result}
\ea
where
\be
A_{\ell,m} \equiv \sqrt{\frac{\ell^2-m^2}{(2\ell-1)(2\ell+1)}}\,. \label{Eq:A_ellm}
\ee
The Remap term requires identities derived in Appendices~\ref{Sect:dipole-X-Ylm}~\&~\ref{Sect:Ylm-derivative}, from which we obtain
\ba
\delta  a_{\ell, m}^\mr{Remap} &= \beta\, \tilde{\bar{N}}(z) \int_{\mathbb{S}_2} \dd^2 \bn \, \dg(z,\bn)  \left[ - (\ell+1) \, A_{\ell,m} \ Y^*_{\ell-1,m}(\bn) + \ell \, A_{\ell+1,m} \ Y^*_{\ell+1,m}(\bn)  \right]  \\
& \quad + 2\beta\, \tilde{\bar{N}}(z) \int_{\mathbb{S}_2} \dd^2 \bn \, \dg(z,\bn)  \left[ A_{\ell,m} \ Y^*_{\ell-1,m}(\bn) + A_{\ell+1,m} \ Y^*_{\ell+1,m}(\bn)  \right] \nonumber\\
&= \beta \left[ -(\ell-1) \, A_{\ell,m} \ \ta_{\ell-1,m} + (\ell+2) \, A_{\ell+1,m} \ \ta_{\ell+1,m}  \right]\,.\nonumber
\ea
In summary, the observed harmonic coefficient relate to the unboosted harmonic coefficients $\ta_{\ell, m}$ by
\ba
a_{\ell, m} &= \ta_{\ell,m} + \frac{\Dkin}{\sqrt{3}} \ta^\mr{BG}_{0,0}\, \delta^\mr{K}_{\ell1}\,\delta^\mr{K}_{m0}
+ \Fkin \, \left[ A_{\ell,m} \, \ta_{\ell-1,m} + A_{\ell+1,m} \, \ta_{\ell+1,m}\right]
- \beta \, \left[ -(\ell-1) \, A_{\ell,m} \, \ta_{\ell-1,m} + (\ell+2) \, A_{\ell+1,m} \, \ta_{\ell+1,m}\right]\,,
\ea
which may be written alternatively as
\ba
a_{\ell,m} &= \ta_{\ell,m} + \frac{\Dkin}{\sqrt{3}} \ta^\mr{BG}_{0,0}\,\delta^\mr{K}_{\ell1}\,\delta^\mr{K}_{m0}
+ A_{\ell,m} \, (\Fkin + \beta(\ell-1)) \ \ta_{\ell-1,m}
+ A_{\ell+1,m} \, (\Fkin - \beta(\ell+2)) \ \ta_{\ell+1,m}\,. \label{Eq:alm_boost}
\ea
The terms proportional to $\beta$ were previously missed in the literature \cite{Dalang2022}. These terms open up the possibility to measure both, $\beta$ and $\Fkin$, independently. We investigate this possibility in detail in the following.

Eq.~\eqref{Eq:alm_boost} has been computed for a fixed redshift $z$. In practice however, one always averages over redshift bins. Labelling by $i$ the $i$th redshift bin, and denoting by $N_i$ the total number of galaxies in bin $i$ we can write the averaged boosted $a_{\ell, m}$ in that bin
\begin{align}
a^\mr{av}_{\ell,m}(z_i) &\equiv\frac{1}{N_i} \int \dd z\, \, a_{\ell, m}(z) \label{Eq:alm_av}\\
&= \ta^\mr{av}_{\ell,m}(z_i)
- \beta \, \left[ -(\ell-1) \, A_{\ell,m} \, \ta^\mr{av}_{\ell-1,m}(z_i) + (\ell+2) \, A_{\ell+1,m} \, \ta^\mr{av}_{\ell+1,m}(z_i)\right] \nonumber\\
&\quad + \frac{1}{N_i}\int \dd z \ \Fkin(z) \, \left[ A_{\ell,m} \, \ta_{\ell-1,m}(z) + A_{\ell+1,m} \, \ta_{\ell+1,m}(z)\right] + \frac{1}{N_i}\int \dd z \ \Dkin(z)\,\ta^\mr{BG}_{0,0}(z)\delta^K_{\ell,1}\delta^K_{m,0}\, \, .\nonumber
\end{align}
In the following, for simplicity we will assume that $\Fkin(z)$ and $\Dkin(z)$ are constant within the redshift bin, such that we can take them out of the integral over $z$ and evaluate them at the mean redshift $z_i$. We recover therefore Eq.~\eqref{Eq:alm_boost}, but with the $a_{\ell, m}(z)$ replaced by $a_{\ell, m}^\mr{av}(z_i)$. To simplify the notation, we drop the superscript ``av'' in the rest of the derivation.

%%%%%%%%%%%%%%%%%%%%%%%%%%%%%%%%%%%%%%%%%%%%%%%%%%%%%%%%%%%%%%%%%%%

\section{Impact on observable statistics}\label{Sect:stat-impact}

In this section, we study the impact of the boost on the statistical properties of the spherical harmonic coefficients. We first recover the leakage of the monopole into the dipole. Then, we construct an observable which is proportional to the boost for which we express the proportionality factor in terms of the clustering properties of the galaxies.
%%%%%%%%%%%%%%%%%%%%
\subsection{Cosmic dipole}\label{Sect:dipole}

For most cosmological signals, the average on the sky is much larger than the fluctuations around the mean. This hierarchy between the monopole ($\propto \ta^{\mr{BG}}_{0,0}$) and the fluctuations ($\tilde{a}_{\ell,m}$ for $\ell \geq 1$) implies that any leakage of the monopole to the higher order multipoles may be important. The effect of the observer's velocity, as described in the previous section, provides such a leakage from the monopole into the dipole, leading to
\ba
a_{1,0} &=  \ta_{1,0} + \frac{D_\mr{kin}}{\sqrt{3}} \ \ta^\mr{BG}_{0,0}+ \Fkin \, A_{1,0} \, \ta_{0,0} + \beta \, (1-1) \, A_{1,0} \, \ta_{0,0} + \mathcal{O}(\ta_{2,0}\,\beta)\nonumber \\
&\simeq \ta_{1,0} + \frac{D_\mr{kin}}{\sqrt{3}}\ \ta^\mr{BG}_{0,0} \,, \label{eq:a10}
\ea
where we have used that the spherical average of $\dg(z,\bn)$ vanishes, since $\dg$ is customarily defined as the deviation from the measured mean number of galaxies, leading to $\ta_{0,0}=0$ in the first line of Eq.~\eqref{eq:a10}. Furthermore, we have neglected the contribution from the quadrupole $\mathcal{O}(a_{2,0}\, \beta)$, as it depends on $\dg$ and is therefore expected to be smaller than $\ta^\mr{BG}_{0,0}D_\mr{kin}$. Under this assumption, we see that the remapping term does not affect the cosmic dipole, which is therefore only sensitive to $D_\mr{kin}$. 

For a more intuitive interpretation, it helps to write this result in real space. Using Eq.~\eqref{Eq:alm_dip} and assuming $|\ta_{\ell ,m}| \ll \ta^{\rm{BG}}_{0,0}$ $\forall \ell \geq 1$, we obtain
\begin{align}
N(z,\bn) & \simeq \sum_{\ell=0}^{1} \sum_{m=-\ell}^\ell a_{\ell, m} Y_{\ell, m}(\bn) 
= \tilde{\bar{N}}(z) \left[ 1 + D_\mr{kin} \sqrt{\frac{4\pi}{3}} Y_{1,0}(\bn) \right] + \sum_{m=-1}^1 \ta_{1,m} Y_{1 ,m}(\bn)  \\
& \simeq \tilde{\bar{N}}(z) \left[1 + D_\mr{kin} \cos(\theta) \right]\, , \label{Eq:Cosmic_Dipole}
\end{align}
where in the second line, we further assumed that $\sqrt{3} \, |\ta_{1,m}| \ll D_\mr{kin} \, \ta_{0,0}^{\rm{BG}} \sim D_\mr{kin}\tilde{\bar{N}}(z)$. This amounts to neglecting the intrinsic dipole in comparison to the kinematically generated dipole. This assumption may not always be a good assumption, especially at low redshift, where the intrinsic dipole due to density fluctuations may be non-negligible. See for example  \cite{Nadolny:2021hti,Cheng2023} for a study of the intrinsic dipole of radio galaxies and \cite{Sorrenti:2022zat} for the dipole of type 1a Supernovae. Eq.~\eqref{Eq:Cosmic_Dipole} is the usual expression for the kinematic dipole, previously derived in~\cite{Maartens:2017qoa,Nadolny:2021hti,Dalang2021}. Let us now study the impact of the boost on the correlation between nearest multipoles.

%%%%%%%%%%%%%%%%%%%%
\subsection{Correlation between multipoles}

We want to compute the correlations between different multipoles due to the observer velocity. We assume that the galaxy density $\dg(z,\bn)$ is statistically isotropic.
This translates to the 2-point correlation function of the spherical harmonic coefficients obeying
\begin{align}
\mathds{E}\left[ \ta_{\ell,m} \ \ta_{\ell',m'}^* \right] = C_\ell \ \delta^\mr{K}_{\ell \ell'} \, \delta^\mr{K}_{m m'}\,. \label{Eq:defCl}
\end{align}
Note that we use the notation $\mathds{E}$ for expectation values (or ensemble averages), and we reserve brackets for the definition of an inner product later in Section~\ref{Sect:estimators}.
The situation becomes different for the boosted spherical harmonic coefficients $a_{\ell,m}$. Indeed, the boost breaks statistical isotropy by introducing a preferred direction in the sky: the direction of $\bm{v}_o$. Setting $\bm{v}_o \propto \bm{e}_z$, we have fount that $a_{\ell,m}$ depend on $\ta_{\ell\pm 1,m}$, which leads to cross-correlations between different $\ell$'s that  no longer vanish. To measure this departure from isotropy, we introduce the following observable, which depends on $m$ and $\ell$
\ba
O_{\ell,m} \equiv a_{\ell,m} \ a^*_{\ell+1,m}\,. \label{Eq:O_l,m}
\ea
At first order in $\beta$, the statistical average of $O_{\ell, m}$ is given by
\ba
\mathds{E}\left[ O_{\ell,m} \right] &= \mathds{E}\left[ \ta_{\ell,m} \times \left(\delta a_{\ell+1,m}^\mr{F} - \delta a_{\ell+1,m}^\mr{Remap}\right)^* \right] + \mathds{E}\left[ \left( \delta a_{\ell,m}^\mr{F} - \delta a_{\ell,m}^\mr{Remap} \right) \times \ta_{\ell+1,m}^* \right] \nonumber\\
&= \mathds{E}\left[ \ta_{\ell,m} \times \left( \Fkin \, A_{\ell+1,m} \, \ta^*_{\ell,\, m} + \beta \, \ell \, A_{\ell+1,m} \, \ta^*_{\ell,m} \right) \right] + \mathds{E}\left[ \left( \Fkin \, A_{\ell+1,m} \, \ta_{\ell+1,\, m} - \beta (\ell+2) \, A_{\ell+1,m} \ta_{\ell+1,m}\right) \times \ta^*_{\ell+1,m} \right] \nonumber\\
&= (\Fkin + \beta \, \ell) \, A_{\ell+1,m} \, C_{\ell} + (\Fkin - \beta \, (\ell+2) ) \, A_{\ell+1,m} \, C_{\ell+1} \nonumber\\
&= A_{\ell+1,m} \cdot \left[ \Fkin \Big(C_\ell + C_{\ell+1}\Big) - \beta \Big( (\ell+2) \, C_{\ell+1} - \ell \, C_\ell \Big)\right]\,. \label{Eq:Olmaverage}
\ea 
For the statistical analysis in Sec.\,\ref{Sect:estimators} and \ref{Sect:forecast}, we also need the covariance matrix of the $O_{\ell,m}$, which is defined as
\ba
\Cov\left(O_{\ell_1,m_1},O_{\ell_2,m_2}\right) & \equiv \mathds{E}\left[ O_{\ell_1,m_1} \, O^*_{\ell_2,m_2} \right] - \mathds{E}\left[ O_{\ell_1,m_1} \right] \mathds{E}\left[ O^*_{\ell_2,m_2} \right] \,.
\ea
We derive this covariance using the $\fsky$ approximation\footnote{This is a popular approximate treatment of partial sky coverage. In Appendix~\ref{App:QE-and-psky}, we sketch how to treat exactly partial sky. We leave the implementation of this exact treatment for future work, as we expect partial sky coverage to have a minimal impact on the high multipoles where the bulk of our statistical power lies.} and assuming gaussianity for the galaxy counts.\footnote{This sounds like a strong assumption when pushing the analysis to high multipoles. However, we explain in Appendix~\ref{App:revising-assumptions-Gaussian-covariance} why this can be assumed: because non-Gaussian contributions have a negligible impact for our specific target, $\beta$, to the level of precision we can claim with this method.} %footnotes always after the punctation in english
To leading order in $\beta$, it gives
\ba
\Cov\left(O_{\ell_1,m_1},O_{\ell_2,m_2}\right) &= \frac{C_{\ell_1} \, C_{\ell_1+1}}{\fsky} \ \delta^\mr{K}_{\ell_1\ell_2} \, \delta^\mr{K}_{m_1m_2} + \mathcal{O}(\beta)\,. \label{Eq:Covariance_O_lm}
\ea

As expected from statistical isotropy, we see that for a non-boosted observer ($\beta=0$), the statistical average of $O_{\ell, m}$ in Eq.~\eqref{Eq:Olmaverage} vanishes: there are no correlations between different multipoles. The boost, however, breaks statistical isotropy. To intuitively understand why this leads to correlations between different $\ell$'s we can look, for example, at the effect of aberration. Perturbations in the hemisphere in the direction of motion subtend a slightly smaller solid angle than in the opposite hemisphere as a result of aberration of the solid angle in which one counts $N(z,\bn)$. The consequence is that perturbations of different scales, parametrized by the $\ell$ index appear to a moving observer to be correlated. From Eq.~\eqref{Eq:Olmaverage}, we see that the observable $O_{\ell, m}$ contains two contribution: one proportional to $\Fkin$, and another one, the remapping term, directly proportional to $\beta$. Since these two contributions scale differently with $\ell$, as we see from Eq.~\eqref{Eq:Olmaverage}, we can in principle measure them separately by combining different $\ell$'s. This is particularly interesting, since it allows us to measure $\beta$ without needing prior knowledge of the galaxy specificities, parametrised by $x(z)$ and $f_\mr{evol}(z)$ which enter in $\Fkin$, see Eq.\,\eqref{Eq:Dkinz}.\footnote{Note that the parameters that enter in $\Fkin(z)$ are not the same as those that contribute to the integrated dipole $\bar{D}_\mr{kin}$ defined in Eq.~\eqref{Eq:EB}. As demonstrated in~\cite{Dalang2021}, the integrated dipole depends on magnification bias and on the spectral index of the sources, whereas the dipole at fixed redshift $z$, as well as the correlations between different multipoles, both depend on magnification bias and evolution bias.} From Eq.~\eqref{Eq:Olmaverage} we see that our ability to separate the two contributions is directly related to the shape of the $C_\ell$'s. A spectrum with structures will help breaking the degeneracies between the two terms. We shall discuss this in more detail in the next section.

%%%%%%%%%%%%%%%%%%%%%%%%%%%%%%%%%%%%%%%%%%%%%%%%%%%%%%%%%%%%%%%%%%%

\section{Estimators for \texorpdfstring{$\beta$}{beta} and \texorpdfstring{$\Fkin$}{Fkin}}\label{Sect:estimators}

In the previous section, we have introduced observables, the $O_{\ell,m}$'s which have a contribution proportional to $\Fkin$ and another one directly proportional to the observer velocity $\beta$. Since some part of the coefficients in front of each of these contributions are known, such as the $A_{\ell, m}$ via Eq.\,\eqref{Eq:A_ellm}, or can be estimated from the measured galaxy number counts, such as the $C_\ell$'s, we can build estimators for $\beta$ and $\Fkin$ from the observable $O_{\ell,m}$'s. In the first subsection, we detail how much can be achieved in an agnostic single tracer case. We then generalize the estimators to the case of multitracers and redshift tomography and finally, we study the case where we we have some prior knowledge on $\Fkin$.

%%%%%%%%%%%%%%%%%%%%
\subsection{Single tracer case}\label{Sect:estimators-single-tracer}

In this section, we lay out the procedure to estimate $\beta$ and $\Fkin$ separately from the observed $O_{\ell, m}$'s.

We start by rewriting Eq.~\eqref{Eq:Olmaverage} in vector form
\ba
\mathds{E}\left[ \bm{O} \right] = \Fkin \ \bm{\alpha} + \beta \ \bm{\gamma}\,, \label{eq:Exptation_Olm}
\ea
where $\bm{O}$ is the vector collecting all $O_{\ell,m}$, and the same goes for $\bm{\alpha}$ and $\bm{\gamma}$, with
\ba
\alpha_{\ell,m} &= A_{\ell+1,m} \ \left(C_\ell + C_{\ell+1}\right) = \sqrt{\frac{(\ell+1)^2-m^2}{(2\ell+1)(2\ell+3)}} \ \left(C_\ell + C_{\ell+1} \right) \,, \label{eq:alpha_lm}\\
\gamma_{\ell,m} &= -A_{\ell+1,m} \ \left( (\ell+2) \, C_{\ell+1} - \ell \, C_\ell \right) = -\sqrt{\frac{(\ell+1)^2-m^2}{(2\ell+1)(2\ell+3)}} \ \left( (\ell+2) \, C_{\ell+1} - \ell \, C_\ell \right)\,. \label{eq:gamma_lm}
\ea
The coefficients $\alpha_{\ell,m}$ and $\gamma_{\ell,m}$ depend on the power spectra $C_{\ell}$. They can be computed either from measured power spectra or from theoretically modelled power spectra. The advantage of using measured power spectra is that we can include highly non-linear scales, where we do not have a reliable theoretical modelling. On the other hand, the theoretical power spectra have the advantage of not being affected by measurement uncertainties. In Appendix~\ref{App:revising-assumptions-known-Cl}, we show that the optimal solution is to use a theoretical model at low $\ell$, where the measurement uncertainties are large (due to cosmic variance) and perturbation theory is reliable. At large $\ell$ on the other hand, it is better to use the measured power spectra that are very well measured and whose errors give negligible contributions to the variance of the estimators. In the following we thus neglect the spectra errors and consider them fixed.

There is a scenario where we can easily derive optimal (meaning unbiased and with minimal variance) Quadratic Estimators (QEs): when one of the parameters dominate the observable, i.e.
\ba\label{Eq:single-param-QEs}
\mr{either} \qquad \Fkin \ \alpha_{\ell,m} \ll \beta \ \gamma_{\ell,m} \qquad \mr{or} \qquad \Fkin \ \alpha_{\ell,m} \gg \beta \ \gamma_{\ell,m}\,.
\ea
In that case we estimate only one parameter, the dominant one. This a problem that has been solved in the CMB lensing literature, see e.g. \cite{Hu2002}. As shown in Appendix~\ref{App:deriv-QE-singleparam} the optimal QE is
\ba
\mr{either} \qquad \hbeta = \frac{1}{\|\bm{\gamma}\|^2} \lbra \bm{O}, \bm{\gamma} \rbra \qquad \mr{or} \qquad \hF = \frac{1}{\|\bm{\alpha}\|^2} \lbra \bm{O}, \bm{\alpha} \rbra\, ,
\ea
where we defined the inner product
\ba
\lbra \bm{x} , \bm{y} \rbra \equiv \sum_{\ell_1,m_1;\ell_2,m_2} x^{*}_{\ell_1,m_1} \, \left(\Sigma^{-1}\right)_{\ell_1,m_1;\ell_2,m_2} \, y_{\ell_2,m_2} = \sum_{\ell,m} \frac{ x^{*}_{\ell,m} \, y_{\ell,m} }{ C_{\ell} \, C_{\ell+1}} \fsky\,,
\ea
with $\Sigma_{\ell_1,m_1;\ell_2,m_2} \equiv \Cov\left(O_{\ell_1,m_1},O_{\ell_2,m_2}\right)$ and where $||\bm{x} ||^2 = \langle \bm{x},\bm{x} \rangle$.

This scenario, where one parameter dominates, leads to simple estimators but the assumptions are not expected to be verified in practice since $\beta$ and $\Fkin$ are generally of the same order of magnitude, and $\bm{\alpha}$ and $\bm{\gamma}$ are also of the same order magnitude. One can show that the estimators of Eq.~\eqref{Eq:single-param-QEs} can still be used (simultaneously) if the weights are orthogonal in the sense of the inner product, i.e.\
\ba
\mr{if} \qquad \lbra \bm{\alpha}, \bm{\gamma} \rbra = 0\,. \label{Eq:orthogonality-alpha-gamma}
\ea
In that case, both parameters are estimated independently. Unfortunately, condition \eqref{Eq:orthogonality-alpha-gamma} is also not expected to hold in practice. An interesting counter-example is that for $C_{\ell+1} = C_{\ell}$, we have $\gamma_{\ell,m}=\alpha_{\ell,m}$. So if all $C_\ell$'s are equal (e.g. in the shot-noise dominated regime), then $\beta$ and $\Fkin$ are perfectly degenerate. A fortiori they cannot be estimated independently.\\
Instead, both parameters have to be estimated jointly. This requires generalising the derivation of Quadratic Estimators as explained in Appendix~\ref{App:deriv-QE-twoparams}. Applying to the current case, we find that the optimal QEs are the estimators $(\hF,\hbeta)$ that solve the linear system
\ba
\begin{pmatrix} \| \bm{\alpha} \|^2 & \lbra \bm{\alpha} , \bm{\gamma} \rbra \\ \lbra \bm{\alpha} , \bm{\gamma} \rbra & \| \bm{\gamma} \|^2 \end{pmatrix}
\times
\begin{pmatrix} \hF \\ \hbeta \end{pmatrix}
=
\begin{pmatrix} \lbra \bm{O} , \bm{\alpha} \rbra \\ \lbra \bm{O} , \bm{\gamma} \rbra \end{pmatrix} \,.\label{Eq:linear-system-Fkin-beta}
\ea
This system is invertible unless the Cauchy-Schwarz inequality is saturated $\left| \lbra \bm{\alpha} , \bm{\gamma} \rbra \right| = \| \bm{\alpha} \| \ \| \bm{\gamma} \|$. In that case $\bm{\alpha}$ and $\bm{\gamma}$ are collinear, meaning that $\beta$ and $\Fkin$ are degenerate and cannot be measured separately because they have the same impact on the data $\bm{O}$. We examine the degenerate case in more details in Sec.~\ref{App:interesting_limits}.\\
Inverting Eq.~\eqref{Eq:linear-system-Fkin-beta} we obtain 
\ba
\begin{pmatrix} \hF \\ \hbeta \end{pmatrix}
&= \frac{1}{1-r_{\bm{\alpha},\bm{\gamma}}^2}
\begin{pmatrix} \| \bm{\alpha} \|^{-2} & -\frac{r_{\bm{\alpha},\bm{\gamma}}}{\|\bm{\alpha}\| \, \|\bm{\gamma}\|} \\ -\frac{r_{\bm{\alpha},\bm{\gamma}}}{\|\bm{\alpha}\| \, \|\bm{\gamma}\|} & \| \bm{\gamma} \|^{-2} \end{pmatrix}
\times
\begin{pmatrix} \lbra \bm{O} , \bm{\alpha} \rbra \\ \lbra \bm{O} , \bm{\gamma} \rbra \end{pmatrix} \,,\label{Eq:QEs-Dkin-beta}
\ea
where we have introduced the correlation coefficient
\ba
r_{\bm{\alpha},\bm{\gamma}} \equiv \frac{\lbra \bm{\alpha} , \bm{\gamma} \rbra }{\| \bm{\alpha} \| \ \| \bm{\gamma} \|}\,.\label{Eq:Correlation_Coefficient}
\ea
From Eq.~\eqref{Eq:QEs-Dkin-beta}, we see that $\hF$ and $\hbeta$ can be directly inferred from the measurements of the observable $O_{\ell, m}$ and the modelling of the coefficients $\alpha_{\ell, m}$ and $\gamma_{\ell, m}$.

To determine how well $\hF$ and $\hbeta$ can be measured, we need to compute the variance of these estimators, as well as their covariance. Those depend on the variance and covariance of $\lbra \bm{O} , \bm{\alpha} \rbra$ and $\lbra \bm{O} , \bm{\gamma} \rbra$. At zeroth order in $\beta$, we obtain 
\ba
\mr{Var}\left(\lbra \bm{O} , \bm{\alpha} \rbra\right) &= \| \bm{\alpha} \|^2 \,, \label{eq:covO}\\
\mr{Var}\left(\lbra \bm{O} , \bm{\gamma} \rbra\right) &= \| \bm{\gamma} \|^2 \,,\\
\mr{Cov}\left( \lbra \bm{O} , \bm{\alpha} \rbra , \lbra \bm{O} , \bm{\gamma} \rbra \right) &= \lbra \bm{\alpha} , \bm{\gamma} \rbra \,.
\ea
With this, after some algebra, we obtain for the variance and covariance of $\hF$ and $\hbeta$
\ba
\mr{Var}\left(\hF\right) &= \frac{1}{\| \bm{\alpha} \|^2 (1-r_{\bm{\alpha},\bm{\gamma}}^2)} \,, \label{Eq:Var_Dkin}\\
\mr{Var}\left(\hbeta\right) &=  \frac{1}{\| \bm{\gamma} \|^2 (1-r_{\bm{\alpha},\bm{\gamma}}^2)}\,, \label{Eq:Var_beta}\\
\Cov\left(\hF, \hbeta\right) &= -\frac{r_{\bm{\alpha},\bm{\gamma}}}{\| \bm{\alpha} \| \, \| \bm{\gamma} \| (1-r_{\bm{\alpha},\bm{\gamma}}^2)}\,.
\ea
The Pearson correlation coefficient for $\hF$ and $\hbeta$ is directly related to $r_{\bm{\alpha},\bm{\gamma}} $ 
\ba
\frac{\Cov\left(\hF, \hbeta\right)}{\sqrt{\mr{Var}\left(\hF\right) \times \mr{Var}\left(\hbeta\right)}} &= 
  - r_{\bm{\alpha},\bm{\gamma}} = - \frac{\lbra \bm{\alpha} , \bm{\gamma} \rbra}{\| \bm{\alpha} \| \| \bm{\gamma} \|)}\,.
\ea
From this, we see that the degeneracy between $\hF$ and $\hbeta$ is directly linked to the degeneracy between the fitting templates $\bm{\alpha}$ and $\bm{\gamma}$, which increases the error bars on $\hF$ and $\hbeta$ through the denominators of Eq.~\eqref{Eq:Var_Dkin}-\eqref{Eq:Var_beta}.

In this section, we have derived analytical expressions for optimal QEs to estimate $\beta$ and $\Fkin$. We now want to gain analytical intuition on the potential of these estimators and their meaning.

\subsection{Simple examples for the estimator and the signal}\label{App:interesting_limits}

In this section we consider some simplified cases in order to gain intuition on the potential to measure $\beta$ and $\Fkin$ and break their degeneracy. 

\subsubsection{One parameter case}

From previous considerations, if $\hF \cdot \alpha_{\ell, m} \gg \hbeta\cdot\gamma_{\ell, m}$ (i.e.\,if the $\Fkin$ effect dominates the equations)  or if $|r_{\bm{\alpha},\bm{\gamma}}| \ll 1$ (i.e.\,the two parameters are decorrelated), then the $\hF$ estimator reduces to 
\begin{equation}
\hF = \frac{1}{\| \bm{\alpha} \|^2} \lbra \bm{O} , \bm{\alpha} \rbra \, ,  
\end{equation}
and its variance is given by $\mr{Var}\left(\hF\right)^{-1} = \| \bm{\alpha} \|^2$. We can input the analytical expressions for $\bm{\alpha}$ and $\Sigma$, to obtain an analytical estimate for this variance in terms of $\ell_\mr{max}$
\be
\frac{\mr{Var}\left(\hF\right)^{-1}}{\fsky} = \sum_{\ell,m}  A_{\ell+1,m}^2  \frac{\left(C_\ell + C_{\ell+1} \right)^2}{ C_{\ell} \, C_{\ell+1}}  \simeq \sum_{\ell,m} A_{\ell+1,m}^2 \times 4 = \frac{4}{3} \sum_{\ell=2}^{\ell_\mr{max}} (\ell+1) = \frac{2}{3} \left(\lmax^2+3\ell_\mr{max} - 4\right)\, , \label{Eq:Var_Dkin_estimator_singleparam}
\ee
where we assumed $C_\ell \simeq C_{\ell +1}$ and used Eq.~\eqref{Eq:sum-A_(l+1,m)} in Appendix~\ref{app:spherical} to compute $\sum_{m} A_{\ell+1,m}^2$.
This means that $\Fkin$ can be measured with an absolute precision $\sim \sqrt{3/2\fsky}\, \lmax^{-1}  $ in this single parameter case. Since $\Fkin \propto \beta\sim 10^{-3}$ we see that we need $\ell_\mr{max}\sim 10,000$ to reach a relative precision of 10 percent with coming large surveys.

In the opposite limit, when $\hF \cdot \alpha_{\ell, m} \ll \hbeta \cdot \gamma_{\ell, m}$ or if $|r_{\bm{\alpha},\bm{\gamma}}| \ll 1$, the $\hbeta$ estimator reduces to 
\begin{align}
\hbeta =  \frac{1}{| \bm{\gamma} \|^{2}}\lbra \bm{O} , \bm{\gamma} \rbra\, ,   
\end{align}
and its variance follows
\ba
\frac{\mr{Var}\left(\hbeta\right)^{-1}}{\fsky} &= \| \bm{\gamma} \|^2 = \sum_{\ell,m}  A_{\ell+1,m}^2 \frac{\left( (\ell+2) \, C_{\ell+1} - \ell \, C_\ell \right)^2}{C_{\ell} \, C_{\ell+1}} = \sum_{\ell}  \frac{\ell+1}{3} \frac{\left( (\ell+2) \, C_{\ell+1} - \ell \, C_\ell \right)^2}{C_{\ell} \,C_{\ell+1}}\,.\label{Eq:Var_beta_estimator_singleparam}
\ea
This can be simplified under two conditions: $C_\ell \simeq C_{\ell +1}$, but also $\ell|C_\ell- C_{\ell +1}| \ll C_{\ell +1}$, which is a stronger condition. In this case
\ba
\frac{\mr{Var}\left(\hbeta\right)^{-1}}{\fsky} &\simeq \sum_{\ell,m} A_{\ell+1,m}^2 \times 4 = \frac{4}{3} \sum_{\ell=2}^{\ell_\mr{max}} (\ell+1) = \frac{2}{3} \left(\lmax^2+3\ell_\mr{max} - 4\right) \,.
\ea
If the conditions are not satisfied (which is usually the case), we have to use a numerical computation of Eq.~\eqref{Eq:Var_beta_estimator_singleparam}. In this case, the scaling of the variance with $\ell_\mr{max}$ is not directly straightforward as it depends on the scaling of $C_{\ell}$ with $\ell$. We will see in Sec.~\ref{Sect:forecast} that for a \Euclid-like survey, the variance of $\hbeta$ can become significantly larger than that of $\hF$, due to the partial cancellation of $(\ell+2) \, C_{\ell+1}$ with $\ell \, C_\ell$ in Eq.~\eqref{Eq:Var_beta_estimator_singleparam}.

Eqs.~\eqref{Eq:Var_Dkin_estimator_singleparam} and~\eqref{Eq:Var_beta_estimator_singleparam} hold in the case where one of the term dominates the signal (or when the two parameters are decorrelated). In the general case, the variance of each of the estimator is boosted by the factor $(1-r_{\bm{\alpha},\bm{\gamma}}^2)^{-1}$. In other words, if we note
\ba
M \equiv \begin{pmatrix} \| \bm{\alpha} \|^2 & \lbra \bm{\alpha} , \bm{\gamma} \rbra \\ \lbra \bm{\alpha} , \bm{\gamma} \rbra & \| \bm{\gamma} \|^2 \end{pmatrix}
\ea
we have that $\left(M_{i,i}\right)^{-1}$ is the unmarginalised (or single parameter) error bar, while $\left(M^{-1}\right)_{i,i}$ is the marginalised error bar.\\
We now turn to simple cases where one can estimate the parameter degeneracy and its meaning.

\subsubsection{Power law angular power spectrum}

As a second example we consider a simple case for the behaviour of the signal, assuming that the angular power spectrum scales as a power law, i.e.\,$C_\ell \propto  1/\ell^n$. Recall the expectation value of the observable $O_{\ell,m}$,
\ba
\mathds{E}\left[ O_{\ell,m} \right] &= \alpha_{\ell,m} \ \Fkin + \gamma_{\ell,m} \ \beta  \\
\alpha_{\ell,m} &= A_{\ell+1,m} \ \left(C_\ell + C_{\ell+1}\right) \\
\gamma_{\ell,m} &= -A_{\ell+1,m} \ \underbrace{\left( (\ell+2) \, C_{\ell+1} - \ell \, C_\ell \right)}_{= 2\frac{\dd \, \ell(\ell+1) \, C_\ell}{\dd \, \ell(\ell+1)} \, \simeq \frac{1}{\ell} \frac{\dd \, \ell^2 \, C_\ell}{\dd \, \ell}}
\ea
For a power law behaviour, we see that at high $\ell$ we have $\alpha_{\ell,m} \simeq -\gamma_{\ell,m} \frac{2}{2-n}$ and thus the $\beta$ and $\Fkin$ parameters are essentially degenerate. Thus one can only measure the combination $G=\Fkin-\beta \, \frac{2-n}{2}$.
Using Eq.~\eqref{Eq:Var_beta_estimator_singleparam} and a derivation similar to the one presented in Eq.~\eqref{Eq:Var_Dkin_estimator_singleparam}, one finds that $\mr{Var}\left(\hat{G}\right)^{-1} \!\!\!\sim \frac{2}{3} \, \lmax^2$. 

Three particular values of $n$ are interesting for us. For a typical galaxy sample, at large $\ell$ but in the regime where the clustering signal still dominates over shot-noise, we have $n\simeq 1$, i.e.\,$\ell \times C_\ell \approx \mr{cst}$. In that case, one can measure the combination $\Fkin-\beta/2$, with an absolute precision $\sim \sqrt{\sfrac{3}{2}}/\lmax$. This is in agreement with the signal to noise ratio presented in \cite{Dalang2022} (which only considered the $\bar{D}_\mr{kin}$ contribution). On the other hand, in the shot-noise dominated regime, we have $C_\ell = \mr{cst}$, i.e.\, $n=0$. In that case, the $\hF$ and $\hbeta$ estimators are perfectly degenerate ($\alpha_{\ell,m}=-\gamma_{\ell,m}$) and one can measure the combination $\hF-\hbeta$. This shows that the noise-dominated part of the spectrum is interesting and should not be thrown away since it allows us to measure this specific combination. In particular, combining the clustering part of the spectrum, where $n=1$, with the shot-noise dominated part, where $n=0$, helps us to break the degeneracy between $\hF$ and $\hbeta$. Finally, the case $n=2$ is also interesting: it corresponds asymptotically to a scale-invariant spectrum with $\ell (\ell+1) C_{\ell}=\mr{cst}$. In that case, the remapping term $\beta$ does not contribute to the signal, and one can only measure $\hF$. This is not surprising: since the remapping term redistributes in an anisotropic way fluctuations with different sizes, it has no impact if the fluctuations of all sizes have the same amplitude.

With the examples above, we see that in several simple cases, $\hF$ and $\hbeta$ are degenerate. To lift this degeneracy, it is important that the $C_\ell$'s exhibit some non-trivial structure, like e.g.\ acoustic peaks, or that they change from one power law to another. In the next section, we generalise the approach to a multi-tracer analysis, or similarly an analysis with multiple redshift bins, and we show that in this case the degeneracy between $\hF$ and $\hbeta$ can be significantly reduced.

%%%%%%%%%%%%%%%%%%%%
\subsection{Generalisation to multi-tracer or multiple redshift bins}\label{Sect:estimators-multi-tracer}

Section~\ref{Sect:estimators-single-tracer} dealt with the case of a single tracer in a single redshift bin, for which there was only one $\Fkin$ to estimate. Modern surveys however observe the galaxy distribution in sufficiently large redshift intervals that several bins are necessary. Since $\Fkin = \Fkin(z)$ is a function of redshift (due to the redshift evolution of $r(z),\HH(z)$ as well as the magnification bias $x(z)$ and the evolution bias $f_\mr{evol}(z)$, see Eq.~\eqref{Eq:Fkin}), one may want to estimate it in each redshift bin. Note that this helps to disentangle the intrinsic from the kinematic dipole in the number counts, as shown in \cite{Nadolny:2021hti}. There may also be different surveys observing different sources (spectroscopic galaxies, photometric galaxies, LRGs, H$\alpha$s, \dots ) for which the quantities $\Fkin(z)$ defined in Eq.\,\eqref{Eq:Fkin} differ because the evolution or magnification biases are population-dependent. In contrast, the remapping contribution proportional to $\beta$ is the same for all the samples and all redshifts.

In the following, we consider $n_s$ different galaxy samples. Those can either be different redshift bins, or different tracers, or both.  
We call $\Fkin^j$ with an index $j\in\{1,\dots,n_s\}$ the coefficient $\Fkin$ corresponding to the sample $j$. We have therefore $n_s+1$ quantities to estimate: $\beta$ and $\Fkin^j$ for $j \in \{1,\cdots,n_s\}$. The observables $O_{\ell,m}$  introduced in Eq.\,\eqref{Eq:O_l,m} now depend on a pair of probes $(j_1,j_2)$ :
\ba
O_{\ell,m}^{(j_1,j_2)} \equiv a_{\ell,m}(j_1) \ a_{\ell+1,m}(j_2)^*\,. 
\ea
Some algebra gives the mean and covariance of the observables
\ba
\mathds{E}\left[ O_{\ell,m}^{(j_1,j_2)} \right] &= A_{\ell+1,m} \times \Big[ \Fkin^{(j_2)} \, C_\ell^{(j_1,j_2)} + \Fkin^{(j_1)} \, C_{\ell+1}^{(j_1,j_2)} - \beta \ \left( (\ell+2) \, C_{\ell+1}^{(j_1,j_2)} - \ell \, C_\ell^{(j_1,j_2)} \right) \Big] \,, \label{Eq:meanO_multi}\\
\Cov\left(O_{\ell,m}^{(j_1,j_2)},O_{\ell',m'}^{(j_3,j_4)}\right) &= \frac{C_{\ell}^{(j_1,j_3)} \, C_{\ell+1}^{(j_2,j_4)}}{\fsky} \ \delta^K_{\ell\ell'} \, \delta^K_{m m'} + \mathcal{O}(\beta)\,.
\ea
Note that while $\mathds{E}\left[ O_{\ell,m}^{(j_1,j_2)} \right] $ is real, the measured $O_{\ell,m}$ will in general be a complex number.
The covariance is again diagonal in $(\ell, m)$, but in general, not in $(j_1,j_2)$, since different redshift bins can be correlated, as well as different populations of galaxies. 
It is convenient to write the mean in Eq.~\eqref{Eq:meanO_multi} as the Euclidean scalar product between two vectors
\ba
\mathds{E}\left[ O_{\ell,m}^{(j_1,j_2)} \right] &=  \bm{f}_{\ell,m}^{(j_1,j_2)} \cdot \bm{\theta}\,,
\ea
where $\bm{\theta} \in \mathbb{R}^{n_s+1}$ is the vector of parameters to estimate and for each values of $\ell, m, j_1, j_2$, $\bm{f}_{\ell,m}^{(j_1,j_2)} $ is an auxiliary vector of $n_s+1$ elements. They are explicitly given by
\ba
\bm{\theta}_0 &= \beta\,, & \qquad \left(\bm{f}_{\ell,m}^{(j_1,j_2)}\right)_0 &= A_{\ell+1,m} \times \Big[ -(\ell+2) \, C_{\ell+1}^{(j_1,j_2)} + \ell \, C_\ell^{(j_1,j_2)} \Big]\,, & \\
\bm{\theta}_j &= \Fkin^{(j)}\,, & \qquad \left(\bm{f}_{\ell,m}^{(j_1,j_2)}\right)_j &= A_{\ell+1,m} \times \Big[ C_\ell^{(j_1,j_2)} \delta^K_{j,j_2} + C_{\ell+1}^{(j_1,j_2)} \delta^K_{j,j_1} \Big]\,, & \quad \forall j \in \{1, \dots, n_s\} \,.
\ea
In the following, we note the index of the vector of parameters with a Greek index, e.g.\ $\mu, \nu=0, \cdots,n_s$. We consider objects like $\bm{f}$ as a vector whose elements are the $\left(\bm{f}_{\ell,m}^{(j_1,j_2)}\right)_\mu$. 
In analogy to tensorial calculations, we shall sometimes act with linear algebra operators on the $\mu$ index and sometimes on the $\!\!\phantom{f}_{\ell,m}^{(j_1,j_2)}$ indices, where the context and dimensionality of the operators should be sufficient to understand unambiguously what is meant.\\
As an application, we define a new inner product between objects having both $(\ell,m)$ and $(j_1,j_2)$ indices
\ba
\lbra \bm{x} , \bm{y} \rbra &= \sum_{\ell,m} \left(\sum_{j_1,j_2,j_3,j_4} \left(x_{\ell,m}^{(j_1,j_2)}\right)^* \ \Cov\left(O^{(j_1,j_2)}_{\ell,m},O^{(j_3,j_4)}_{\ell,m}\right)^{-1} \ y_{\ell,m}^{(j_3,j_4)} \right) \\
&= \left(x^A\right)^* \left(\Sigma^{-1}\right)_{AB} y^B\, , \label{Eq:inner_multi}
\ea
where in the second line we used Einstein summation convention and the definition
\be
\Sigma^{AB} = \Cov(O^A,O^B).
\ee
Using this inner product we can define a vector of size $n_s+1$ that we note $\lbra \bm{O}, \bm{f} \rbra$ and a matrix of shape $(n_s+1) \times (n_s+1)$ that we note $\lbra \bm{f}, \bm{f} \rbra$:
\ba
\lbra \bm{O}, \bm{f} \rbra_\mu \equiv \lbra \bm{O}, \bm{f}_\mu \rbra = \left(O^A\right)^* \left(\Sigma^{-1}\right)_{AB} f^B_\mu\, , \\
\lbra \bm{f}, \bm{f} \rbra_{\mu,\nu} \equiv \lbra \bm{f}_\mu, \bm{f}_\nu \rbra = \left(f^A_\mu\right)^* \left(\Sigma^{-1}\right)_{AB} f^B_\nu\, .
\ea
Using this definitions and the derivation of multi-parameters QEs from Appendix~\ref{App:deriv-QE-multiparams}, we find that optimal QEs for the $\bm{\theta}$ components is given by the solution to the following linear system:
\ba
\lbra \bm{f}, \bm{f} \rbra \cdot \bm{\hat{\theta}} = \lbra \bm{O}, \bm{f} \rbra\,.\label{Eq:syseq_multi}
\ea
The system is invertible unless there is a perfect degeneracy between some parameters, which we found not to happen in practice. Inverting the system leads to the estimators $\bm{\hat{\theta}} = \lbra \bm{f}, \bm{f} \rbra^{-1} \cdot \lbra \bm{O}, \bm{f} \rbra$. In the two parameters case of Sec.~\ref{Sect:estimators-single-tracer}, we could invert the matrix $\lbra \bm{f}, \bm{f} \rbra$ analytically. This is no longer possible in the generalised case, although there is no obstacle to numerical inversion in practice.

We now compute the covariances of the $n_s + 1$ estimators. To do so, we start by writing the components of the estimator $\hat{\theta}^{\,\mu}$
\be
\hat\theta^{\,\mu} = \left(\lbra \bm{f} , \bm{f} \rbra^{-1}\right)^{\mu\nu} \left(O^{A}\right)^* \left(\Sigma^{-1}\right)_{AB} f^B_\nu \,.
\ee
Using linearity, the covariance between the different estimators $\hat\theta^{\,\mu}$ becomes
\ba
\Cov\left(\hat\theta^{\,\mu_1},\hat\theta^{\,\mu_2}\right) =& ~ \left(\lbra \bm{f} , \bm{f} \rbra^{-1}\right)^{\,\mu_1\nu_1} \Sigma^{AC} \left(\Sigma^{-1}\right)_{AB} f^B_{\nu_1}  \left(\lbra \bm{f} , \bm{f} \rbra^{-1}\right)^{\,\mu_2\nu_2} \left(\Sigma^{-1}\right)_{CD} \left(f^D_{\nu_2}\right)^* \nonumber\\
=&  ~\delta^C_B \left(\Sigma^{-1}\right)_{CD} f^B_{\nu_1} \left(f^D_{\nu_2}\right)^* \left(\lbra \bm{f} , \bm{f} \rbra^{-1}\right)^{\,\mu_1\nu_1} \left(\lbra \bm{f} , \bm{f} \rbra^{-1}\right)^{\,\mu_2\nu_2} \nonumber
\\
=& ~ \lbra \bm{f} , \bm{f} \rbra_{\nu_1\nu_2}\left(\lbra \bm{f} , \bm{f} \rbra^{-1}\right)^{\,\mu_1\nu_1}\left(\lbra \bm{f} , \bm{f} \rbra^{-1}\right)^{\,\mu_2\nu_2}\nonumber\\
= &~ \left(\lbra \bm{f} , \bm{f} \rbra^{-1}\right)^{\,\mu_1\mu_2}\,, 
\ea
where we have also used the symmetry of the covariance matrix $\Sigma^{AB}$ and of the matrix $\lbra \bm{f} , \bm{f} \rbra$.\\
In particular the variance of the estimator $\hbeta$ is $\left(\lbra \bm{f}, \bm{f} \rbra^{-1}\right)^{00}$. 

Thus, to forecast the uncertainties on the parameters of interest, we need to compute the inverse of the symmetric matrix $\lbra \bm{f}, \bm{f} \rbra$. This calculation can be slightly simplified by performing analytically the sum over $m$. We use that $\Sigma^{AB}$ is diagonal in $\ell$ and $m$ and independent of $m$, and  we denote by $\bm{\Sigma}_\ell$ the matrix $\Sigma_{(\ell\,m)\,(\ell\,m)}^{(j_1,j_3),(j_2,j_4)}=C_{\ell}^{(j_1,j_3)} \, C_{\ell+1}^{(j_2,j_4)} / \fsky$. With this, we find
\ba
\lbra \bm{f}, \bm{f} \rbra_{00} &= \sum_\ell \left(- (\ell+2) \, \bar{\bm{C}}_{\ell+1} + \ell \, \bar{\bm{C}}_\ell \right)^T \cdot \bm{\Sigma}_\ell^{-1} \cdot \left(- (\ell+2) \, \bar{\bm{C}}_{\ell+1} + \ell \, \bar{\bm{C}}_\ell \right) \underbrace{\sum_m A_{\ell+1,m}^2}_{=\frac{\ell+1}{3}} \nonumber \\
&= \sum_\ell \frac{\ell+1}{3} \left(- (\ell+2) \, \bar{\bm{C}}_{\ell+1} + \ell \, \bar{\bm{C}}_\ell \right)^T \cdot \bm{\Sigma}_\ell^{-1} \cdot \left(- (\ell+2) \, \bar{\bm{C}}_{\ell+1} + \ell \, \bar{\bm{C}}_\ell \right)\,, & \label{Eq:var_multi_00}\\
\lbra \bm{f}, \bm{f} \rbra_{0j} &= \sum_\ell \frac{\ell+1}{3} \left(- (\ell+2) \, \bar{\bm{C}}_{\ell+1} + \ell \, \bar{\bm{C}}_\ell \right)^T \cdot \bm{\Sigma}_\ell^{-1} \cdot \left( \bar{\bm{C}}_\ell \, \delta^K_{j,j_4}+ \bar{\bm{C}}_{\ell+1} \, \delta^K_{j,j_3} \right)\,, & \forall j\in \{1,\dots,n_s \}\,, \label{Eq:var_multi_0j}\\
\lbra \bm{f}, \bm{f} \rbra_{ij} &= \sum_\ell \frac{\ell+1}{3} \left( \bar{\bm{C}}_\ell \, \delta^K_{i,j_2}+ \bar{\bm{C}}_{\ell+1} \, \delta^K_{i,j_1} \right)^T \cdot \bm{\Sigma}_\ell^{-1} \cdot \left( \bar{\bm{C}}_\ell \, \delta^K_{j,j_4}+ \bar{\bm{C}}_{\ell+1} \, \delta^K_{j,j_3} \right)\,, & \forall i,j \in \{1,\dots, n_s\}\,.\label{Eq:var_multi_ij}
\ea
where $\bar{\bm{C}}_\ell$ is understood as a vector with size $n_s(n_s+1)/2$ containing $C_\ell^{(j_1,j_2)}$ for all pairs of samples and the indices $(j_1,j_2)$ and $(j_3,j_4)$ are also suppressed in the matrix $\bm{\Sigma}_\ell$.\\
We now have all the tools to forecast the uncertainties on $\hbeta$ and $\hF^{(j)}$.

%%%%%%%%%%%%%%%%%%%%%%%%%%%%%%%%%%%%%%%%%%%%%%%%%%%%%%%%%%%%%%%%%%%

\section{Forecast for future surveys}\label{Sect:forecast}

In this section, we forecast how well a \Euclid-like survey may measure $\beta$ and $\Fkin$, assuming a fiducial cosmological model. We first consider the case of a single tracer for a single bin. We then show how the measurement of $\beta$ is improved when we combine 10 redshift bins. Finally, we study the impact of prior knowledge of $\Fkin$ through a direct measurement of the dipole, and most importantly, the impact of prior knowledge on the source evolution and magnification biases. 

We follow the survey specifications of \cite{Euclid-IST} for a photometric \Euclid-like survey. We choose our fiducial cosmology to be the flat $\Lambda$CDM model with parameters from Planck~\cite{Planc2015-DEMG}, namely $(\Omega_\mr{b},\Omega_\mr{m},h,n_s,\sigma_8,\sum m_\nu/[\mr{eV}])=(0.05,0.32,0.67,0.96,0.816,0.06)$. The matter power spectrum $P(k)$ on non-linear scales follows the prescriptions of \cite{Takahashi2012,Bird2012}.  We consider a  photometric mock galaxy survey of 15,000 deg$^2$  survey with a redshift distribution given by
\be
n(z) \propto \left(\frac{z}{z_0}\right)^2 \, \exp\left[-\left(\frac{z}{z_0}\right)^{3/2}\right]\,,
\ee
where $z_0=0.9/\sqrt{2}$. The normalisation is set by requiring a total density of 30 galaxies per $\mr{arcmin}^2$. We consider 10 redshift bins with boundaries defined such that they are equi-populated. Their boundaries are given by
$$z_i=(0.001,0.42,0.56,0.68,0.79,0.90,1.02,1.15,1.32,1.58,2.50).$$
The redshifts of all sources are then perturbed with photometric errors following the specifics from \cite{Euclid-IST}. Galaxy clustering is modelled as a biased tracer of matter with a scale-independent bias $b(z)=\sqrt{1+z}$. The galaxy angular power spectrum is considered on a range of multipoles from $\ell_\mr{min}=10$ to $\ell_\mr{max}=10,000$. Since we do not have measurements for \Euclid{} power spectra yet, we use theoretical spectra at all scales. Concerning the error budget, we do include cosmic variance and shot noise in our error budget but neglect additional instrumental errors.

%%%%%%%%%%%%%%%%%%%%
\subsection{Individual redshift bins}

\begin{figure}[t]
    \begin{center}
        \includegraphics[width=\linewidth]{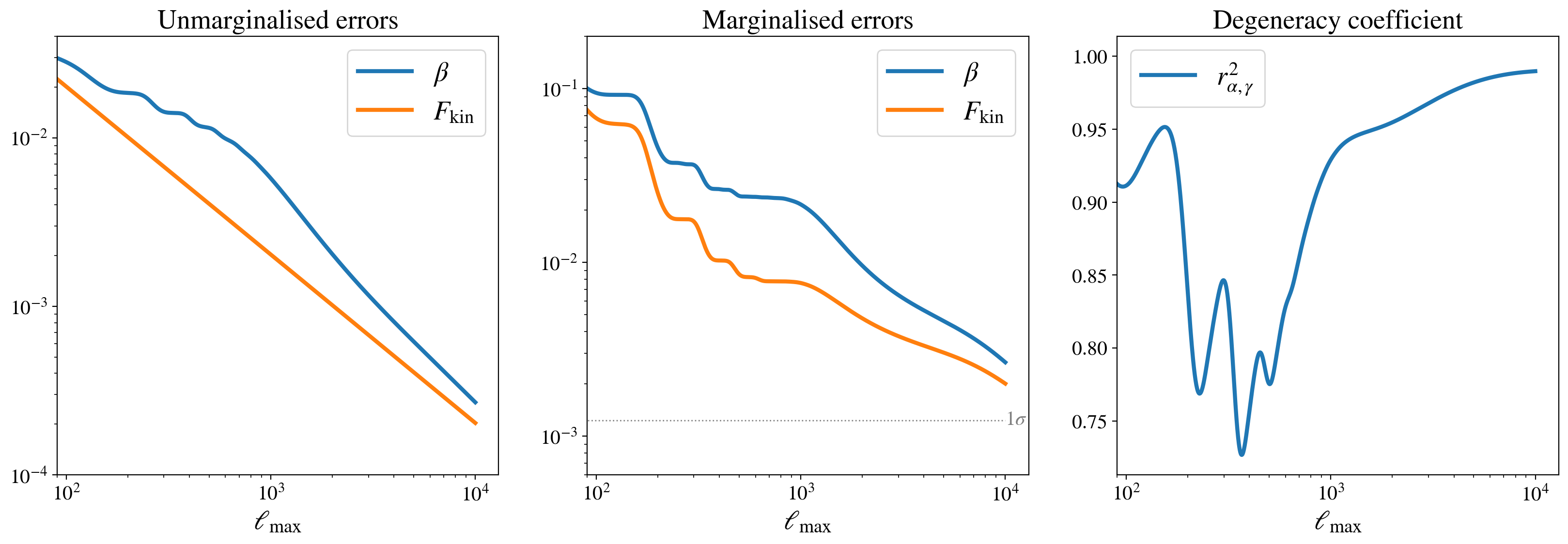}
         \caption{\textit{Left:} forecasted unmarginalised absolute errors on $(\beta,\Fkin)$ from a single tracer analysis using the median redshift bin of a \Euclid-like galaxy survey, as a function of $\lmax$. \textit{Center:} marginalised error. \textit{Right:} degeneracy coefficient between the two parameters. All quantities are plotted as a function of the resolution of the analysis, parameterized by the maximal multipole $\lmax$ included in the data vector. Note that the y-axis of the left and central plots are different, since the marginalisation significantly inflates the errors. In the central plot, the dashed line corresponds to $\beta = 1.23 \cdot 10^{-3}$, which is the expectation from CMB observations. The observer's peculiar velocity falls short of being detectable with such an analysis.}
        \label{Fig:errors_singletracer}
    \end{center}
\end{figure}

We first consider the case of individual redshift bins. In Fig.~\ref{Fig:errors_singletracer} we plot the errors on $\hbeta$ and $\hF$ as a function of the maximal multipole $\lmax$, for the fifth redshift bin of \Euclid{} $z\in[0.79,0.9]$. This is the median redshift bin, as a representative case. In Fig.~\ref{Fig:errors_singletracer}, the left panel shows the unmarginalized errors, which are the inverse square root of the diagonal elements of the $\langle \bm{f},\bm{f}\rangle $ matrix, while the middle panel shows the marginalized errors, which are the diagonal elements of $\langle \bm{f},\bm{f}\rangle^{-1}$. We also plot in the right panel, the correlation coefficient defined in Eq.~\eqref{Eq:Correlation_Coefficient}, which governs the covariance between the estimators $\hbeta$ and $\hF$. Looking at the unmarginalised errors, we see that $\hF$ is better measured than $\hbeta$, especially for $\ell_\mr{max} \lesssim 1000$. This is due to the partial cancellation of the contribution proportional to $\beta$ in Eq.~\eqref{Eq:Olmaverage}, when $(\ell+2)C_{\ell+1}\sim \ell C_\ell$. This in turns increases the variance of $\hbeta$, as we see from Eq.~\eqref{Eq:Var_beta_estimator_singleparam}. In particular, when $n=1$ (clustering signal at large $\ell$), we have $(\ell+2)C_{\ell+1}- \ell C_\ell\sim 2/\ell$, leading to a large variance of $\hbeta$. Adding more multipoles helps since it adds a part of the spectrum where shot-noise is non-negligible. In particular, in the regime where shot-noise dominates ($n=0$) we have $(\ell+2)C_{\ell+1}- \ell C_\ell\sim 2$, which decreases the variance of $\hbeta$.

From the middle panel of Fig.~\ref{Fig:errors_singletracer}, we see that the marginalised errors are significantly larger than the unmarginalised one, by almost a factor 10 for high $\ell_\mr{max}$. This is due to the large degeneracy between $\hbeta$ and $\hF$. The correlation coefficient plotted in the right panel of Fig.~\ref{Fig:errors_singletracer} is indeed close to 1 at large $\ell_\mr{max}$. It is only around the BAO scales, where the $C_\ell$'s have a lot of structure, that the two contributions become less correlated. As seen from Eqs.~\eqref{Eq:Var_Dkin} and~\eqref{Eq:Var_beta}, this strong degeneracy between $\hbeta$ and $\hF$  increases the variance of the estimators and consequently prevents us from detecting a $\beta$ of the order of $10^{-3}$, even if one pushes the analysis to $\lmax=10^{4}$.

\begin{figure}[!ht]
    \begin{center}
        \includegraphics[width=\linewidth]{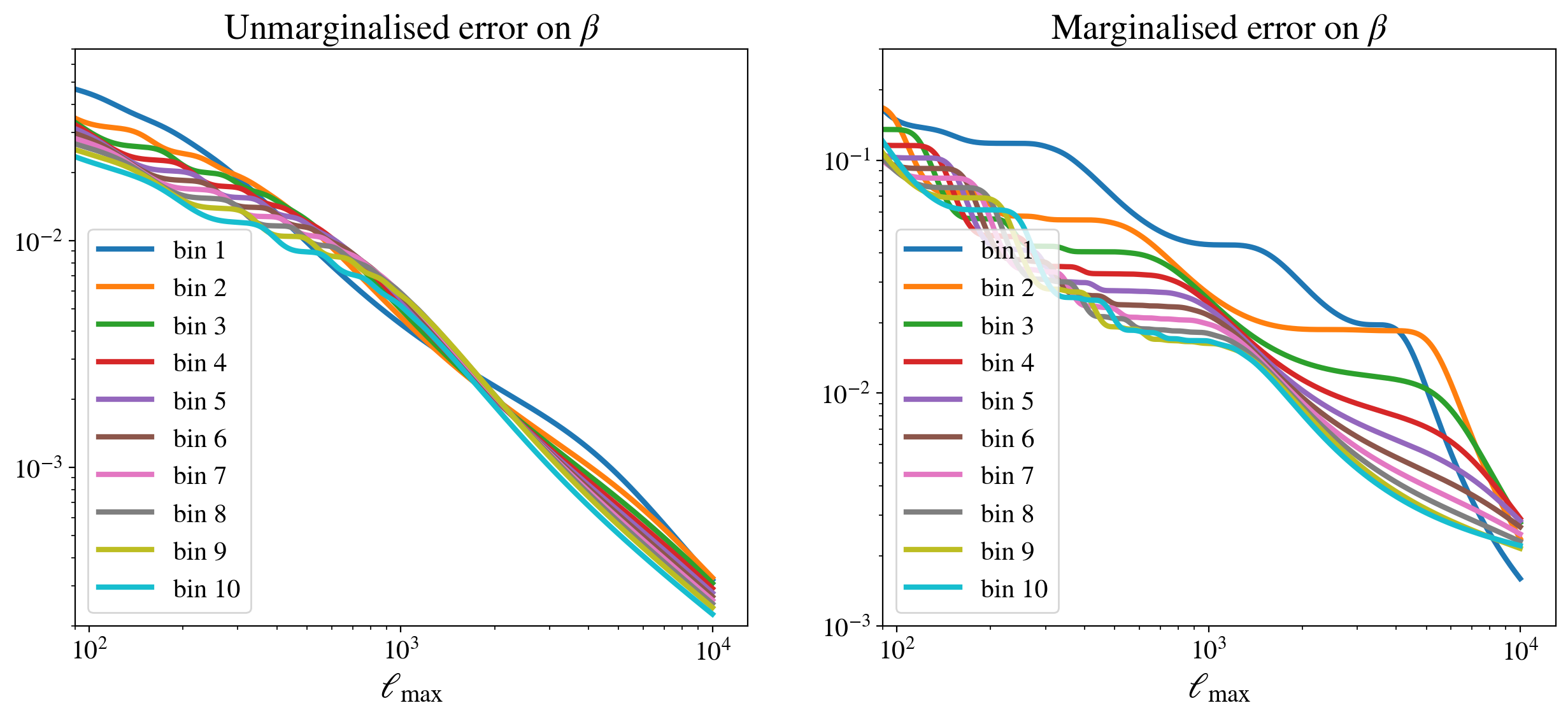}
         \caption{Forecasted absolute error on $\beta$ as a function of $\lmax$, for the ten redshift bins of an \Euclid-like survey, when each bin is analysed separately. We show the unmarginalised error in the left panel and the marginalised errors in the right panel. Again, the $\beta = 1.23 \cdot 10^{-3}$ value derived from the CMB dipole falls short of being detectable.}
        \label{Fig:error_beta_singletracer}
    \end{center}
\end{figure}

In Fig.~\ref{Fig:error_beta_singletracer}, we present the same results for the 10 photometric redshift bins of Euclid, when each of them is analysed separately. More precisely, we use Eqs.~\eqref{Eq:Var_Dkin} and~\eqref{Eq:Var_beta} to determine the variance on $\hF^{(i)} =\hF(z_i)$ and $\hbeta$ for each bin $i$. In the left panel of Fig.~\ref{Fig:error_beta_singletracer} we show the unmarginalised errors on $\hbeta$ from each of the bins, whereas in the right panel we plot the marginalised errors. For large values of $\ell_\mr{max}$, the error on $\hbeta$ decreases when the redshift bin increases. This may seem surprising, since the angular power spectrum decreases at high redshift, where galaxy clustering is smaller than at low redshift. However, shot noise is larger at high redshift, where less galaxies are detected. As discussed before, since shot noise is affected by the remapping term, proportional to $\beta$, it helps us in measuring $\beta$. Even though the errors are smaller at high redshift, we see from the right panel of Fig.~\ref{Fig:error_beta_singletracer} that the degeneracy between $\hbeta$ and $\hF^i$ is still too large to allow for a 1$\sigma$ detection at $\lmax =10,000$.

To conclude, in this section we artificially considered each of the redshift bins individually and found that $\beta$ could reasonably not be detected. However, the remapping contribution proportional to $\beta$ is the same for all redshifts. Therefore, one can hope that combining redshift bins will improve the measurement of $\beta$ and reduce the degeneracies. This is what we turn to in the next section.

\subsection{Combining 10 redshift bins}

\begin{figure}[!ht]
    \begin{center}
        \includegraphics[width=\linewidth]{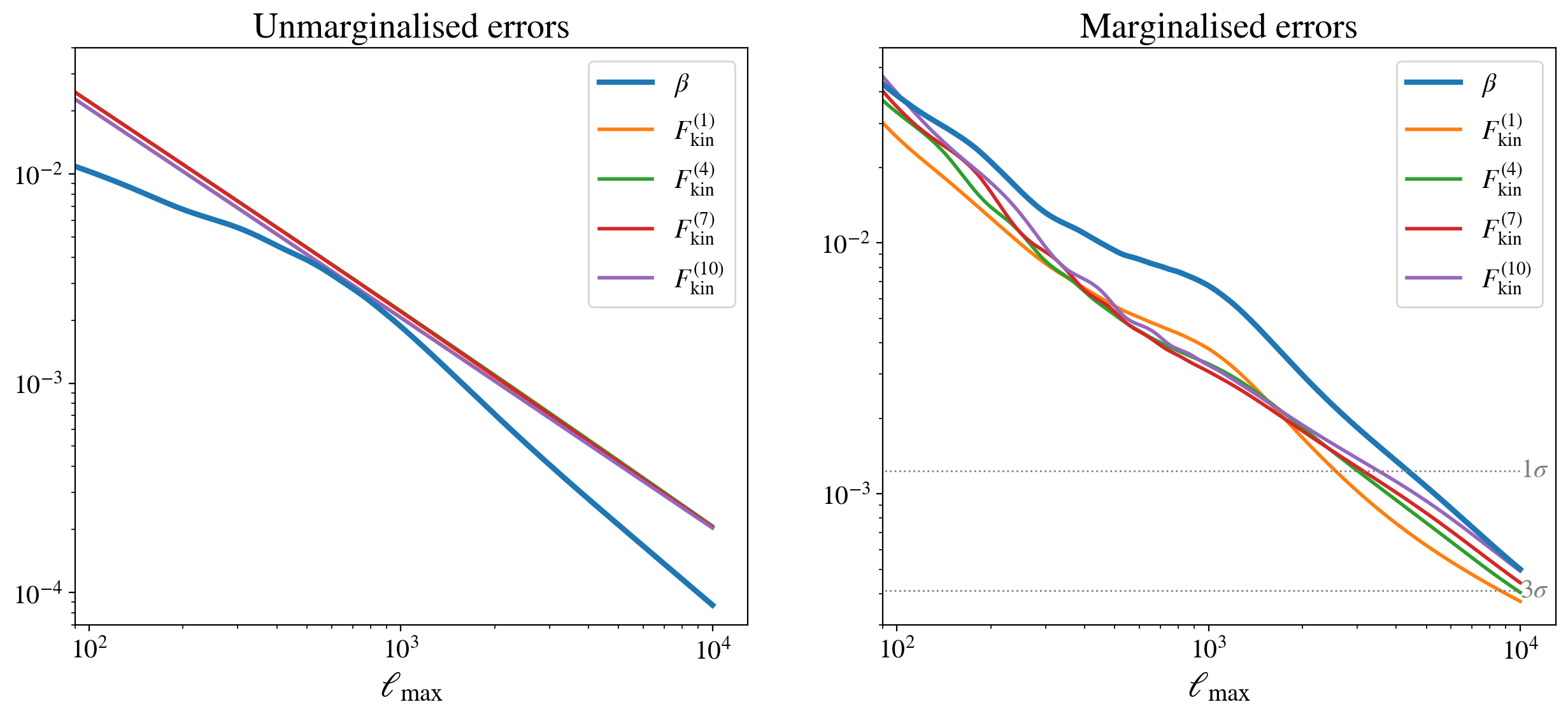}
         \caption{Forecasted absolute errors on $\beta$ and $\Fkin^{(i)}$ as a function of $\lmax$ for a few redshift bins $i\in \{1,4,7,10\}$ for a multitracer analysis using the full power of the ten redshift bins of the \Euclid-like survey. We show the unmarginalised error in the left panel and the marginalised errors in the right panel. The CMB value $\beta=1.23\cdot 10^{-3}$ would be detected at $2.4\sigma$ at $\lmax=10,000$.}
        \label{Fig:errors_multitracer}
    \end{center}
\end{figure}

To perform a proper combined analysis of the 10 redshift bins, we need to use the estimators $\bm{\hat{\theta}}$ obtained from solving Eq.~\eqref{Eq:syseq_multi}, where $\hat{\theta}^0=\hbeta$ and $\hat{\theta}^i=\hF^{(i)}$. The variance and covariance of theses estimators are given in Eqs.~\eqref{Eq:var_multi_00} to~\eqref{Eq:var_multi_ij}. To compute them, we need to account for the covariance of the $C_\ell$'s between different redshift bins. In Fig.~\ref{Fig:errors_multitracer}, we show the unmarginalised (left panel) and marginalised (right panel) errors on $\hbeta$ as well as on $\hF^{(i)}$ for four representative redshift bins. As expected, we see that the unmarginalised error on $\hbeta$ combining the 10 bins are improved by a factor $\sim \!3$ with respect to the single bin case. The improvement on the marginalised error on $\hbeta$ is however more significant: the error decreases by a factor $\sim \!5$. This is due to the fact that combining different redshift bins helps breaking the degeneracy between $\hbeta$ and $\hF^{(i)}$ since the contribution from $\hbeta$ to the observable $O_{\ell, m}$ varies with redshift. As a consequence, the errors on $\hF^{(i)}$ and $\hbeta$ all become smaller than $10^{-3}$ around $\lmax\simeq 10^3$. We therefore obtain a $2.4\sigma$ significance of $\hbeta$ at $\lmax=10,000$ if $\beta=1.23\cdot 10^{-3}$.

%%%%%%%%%%%%%%%%%%%%
\subsection{Adding priors on \texorpdfstring{$\Fkin$}{Fkin}}

The multi-tracer analysis presented in the previous section works reasonably well. We now investigate how the constraints improve if there is prior knowledge of some of the quantities involved. At first, we considered modelling the redshift dependence of the coefficients $\hF(z)$ with 2-4 parameters. However, we found that doing so does not improve the marginalized error on $\hbeta$ by more than a few percent.

Another possibility is to add priors on $\hF^{(i)}$ using measurements of the dipole, i.e.\ the spherical harmonic coefficient $a_{1,0}$ in Eq.~\eqref{eq:a10}.
Measuring the dipole bin per bin provides direct measurements of $\Dkin^{(i)}$, which is related to $\Fkin^{(i)}$ by Eq.~\eqref{Eq:Fkin}. The precision on the measurement of $\Dkin^{(i)}$ is limited by shot noise, which scales as $D_\mr{sn} \simeq 3\sqrt{f_\mr{sky}/N_i}$ 
where $N_i$ is the number of sources in the bin $i$, see~\cite{Nadolny:2021hti}. Current measurements have reached a precision of~10\% using more than a million  quasars~\cite{Secrest:2020has}.
Since the \Euclid{} photometric survey~\cite{Euclid-redbook} is expected to detect more than $10^9$ sources, we will have, in each redshift bin, about 100 million galaxies. We expect therefore the shot-noise uncertainty to be 10 times smaller than in~\cite{Secrest:2020has} in each of the redshift bins, leading to a precision on $\Dkin^{(i)}$ of the order of 1\%.

Note however that the dipole $a_{1,0}$ in Eq.~\eqref{eq:a10} depends not only on $\Dkin^{(i)}$ but also on the intrinsic clustering dipole $\ta_{1,0}$ that may be important at low redshift $z\leq 0.1$~\cite{Tiwari:2016,Nadolny:2021hti} and contaminate the measurements of $\Dkin^{(i)}$ in our lowest redshift bin.
Furthermore, $\Fkin^{(i)}$ contains additional contributions with respect to $\Dkin^{(i)}$ (see Eq.~\eqref{Eq:Fkin}), which are sensitive to the growth rate and the evolution of the galaxy bias. 
To account for these two effects, we enlarge our priors on $\Fkin^{(i)}$, choosing two cases in our analysis: a $10\%$ prior and a $30\%$ prior.

\begin{figure}[!ht]
    \begin{center}
        \includegraphics[width=\linewidth]{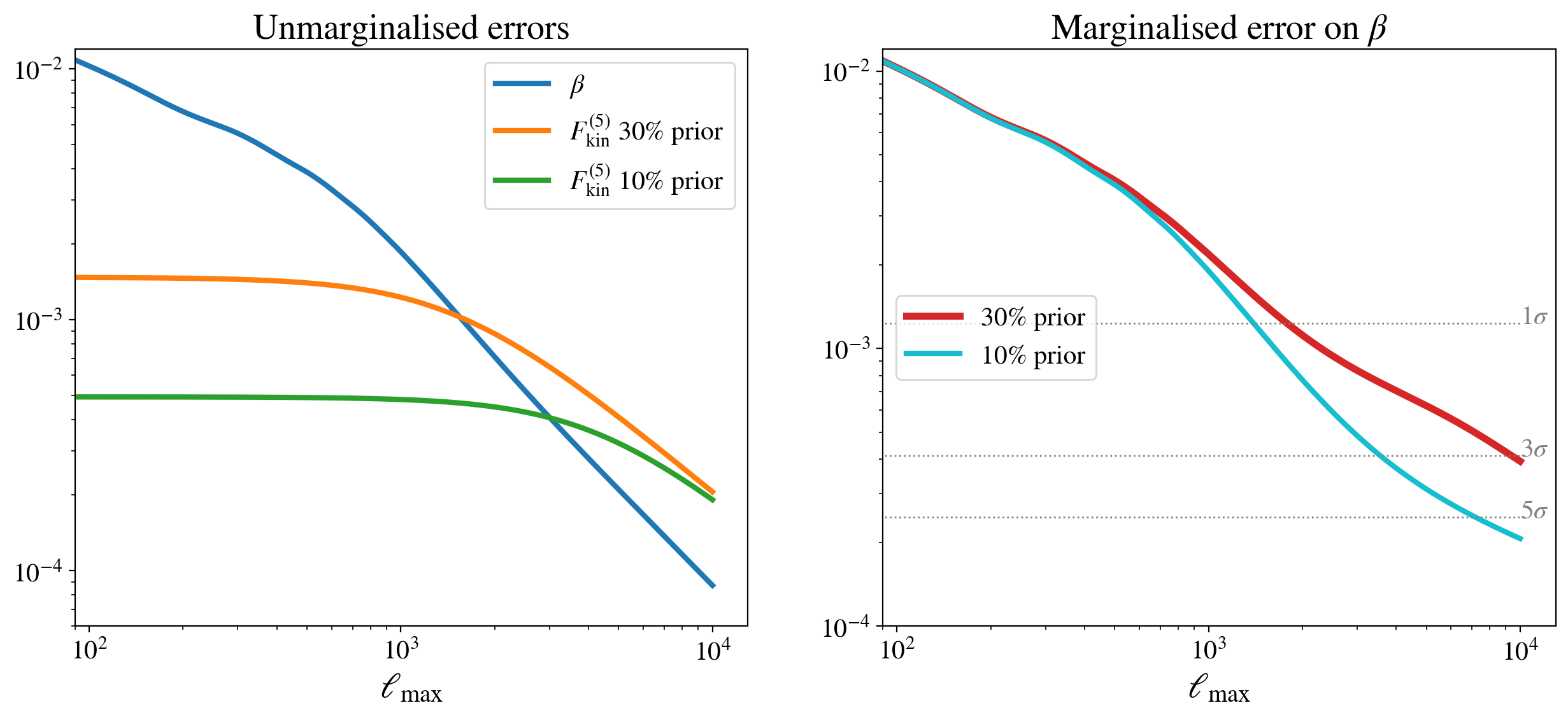}
         \caption{Forecasted absolute errors on $\beta$ and $\Fkin^{(i)}$, using a prior on $\Fkin^{(i)}$. We consider a multitracer analysis with 10 redshift bins of a \Euclid-like survey and plot the errors as a function of $\lmax$. As illustration we choose a fiducial value of $\Fkin^{(i)} = 4\cdot \beta$, with the CMB value $\beta = 1.23 \cdot 10^{-3}$, and we consider two different priors: a $10\%$ prior and a $30\%$ prior. We show the unmarginalised error in the left panel and the marginalised errors in the right panel.
         }
        \label{Fig:errors_multitracer_wprior_Dkin}
    \end{center}
\end{figure}

In Fig.~\ref{Fig:errors_multitracer_wprior_Dkin}, we show the unmarginalised (left panel) and marginalised (right panel) errors on $\hbeta$ and on one of the coefficient $\hF^{(5)}$ combining the 10 redshift bins of \Euclid. For small values of $\ell_\mr{max}$, the unmarginalised error on $\hF^{(5)}$ are dominated by the priors, i.e. our method does not bring information for this parameter. However when increasing $\ell_\mr{max}$, our method gradually decreases the errors down to $2\times 10^{-4}$. From the right panel of Fig.~\ref{Fig:errors_multitracer_wprior_Dkin}, we see that the priors on $\hF^{(5)}$ have a strong impact on the marginalised errors on $\hbeta$. Indeed, a prior knowledge of $\hF^{(5)}$ breaks the strong degeneracy between $\hF^{(5)}$ and $\hbeta$, allowing for a significant detection of $\beta$. We see that with a 30\% prior, $\beta = 1.23\cdot 10^{-3}$ is detectable at 3.1$\sigma$ at $\lmax=10,000$. With a 10\% prior, which is still very reasonable, we reach $3\sigma$ significance at $\lmax=3,569$, $5\sigma$ significance at $\lmax=7,224$, and the maximum significance reached at $\lmax=10,000$ is of $6\sigma$.

\subsection{Adding priors on the magnification bias and evolution bias}

Another way to break the degeneracy between $\hF$ and $\hbeta$ is to model directly the quantities that enter into $\Fkin$, see Eq.~\eqref{Eq:Fkin}. In particular, the magnification bias, evolution bias, as well as the evolution of the galaxy bias with redshift and luminosity threshold $L_*$ can be measured from the population of galaxies. 
The comoving distance, Hubble parameter and growth function, that also affect $\Fkin$, can be calculated assuming a cosmological model. All this information can then be translated into a prior on the coefficients $A_i \equiv \Fkin^{(i)}/\beta$. The error on $A_i$ are expected to be dominated by the errors on the magnification bias, evolution bias and galaxy bias evolution, since the cosmological parameters should be determined with percent precision by future surveys. As shown in~\cite{Wang:2020ibf}, the evolution and magnification bias can be measured with a precision of the order of $10\%$ using $\sim 10^4$ objects. Realistic expectations for \Euclid{} are similar \cite{Euclid:2021rez}.
In our analysis, we take conservative priors of 30\% and 10\% on the coefficients $A_i$, $i\in\{1,\dots, 10\}$. 

\begin{figure}[!ht]
    \begin{center}
        \includegraphics[width=\linewidth]{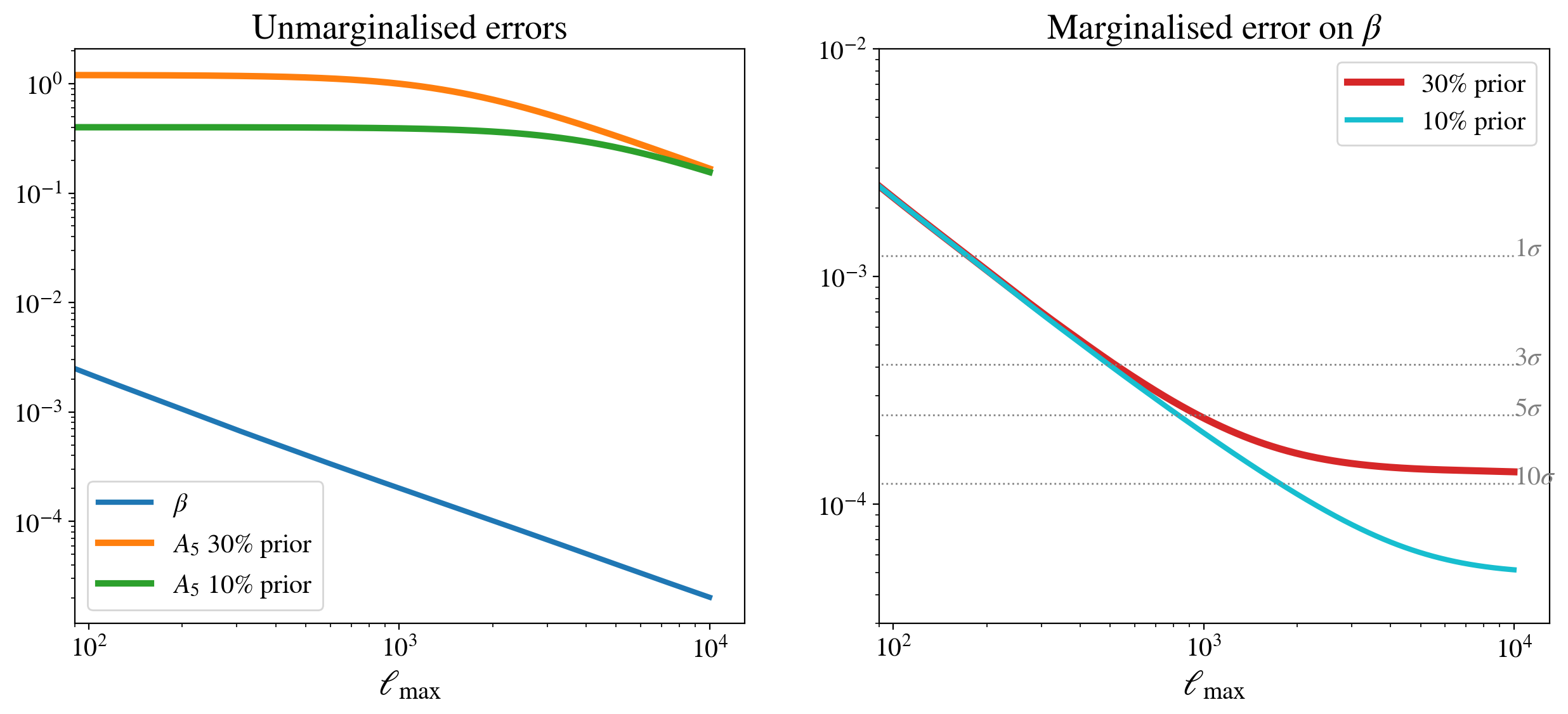}
         \caption{Forecasted absolute errors on $\beta$ and $\Fkin^{(i)}$ as a function of $\lmax$ for a multitracer analysis using the 10 redshift bins of the \Euclid-like survey. Here, we add a prior on the amplitude $A_i = F^{(i)}_\mr{kin}/\beta$, the proportionality factor given in Eq.\,\eqref{Eq:Fkin_gen}. Precise knowledge of this factor is mostly limited by external measurements of the galaxy evolution bias and magnification bias.}
        \label{Fig:errors_multitracer_wprior_Ai}
    \end{center}
\end{figure}

The errors are plotted in Fig.~\ref{Fig:errors_multitracer_wprior_Ai}. We see that in this case, the marginalised error on $\beta$ reaches $10^{-4}$ at $\lmax = 10,000$ for the $30\%$ priors on $A_i$ and $4 \cdot 10^{-5}$ for the $10\%$ priors. In particular, with a 30\% prior, a value $\beta \simeq 1.23 \cdot 10^{-3}$ consistent with the CMB is detectable at $5\sigma$ already for $\lmax=962$. Using all multipoles up to $\lmax=10,000$ we reach a maximum significance of $8.9\sigma$. With a 10\% prior we reach $5\sigma$ significance at $\lmax=832$, $10\sigma$ significance at $\lmax=1,769$, and a maximum significance of $24\sigma$ at $\lmax=10,000$. This corresponds to a relative precision of $4.1\%$ in the measurement of $\beta$. The strong improvement brought by the priors on $A_i$ is due to the fact that in this case, not only $\hbeta$ is used to measure $\beta$, but the 10 values of $\hF^{(i)}$ are also used, since we now use the fact that $\hF^{(i)}$ are proportional to $\beta$ with coefficients $A_i$ that we model with some imprecision. This analysis shows therefore that a survey like \Euclid{} should allow us to determine the observer velocity using correlations between $a_{\ell, m}$, with a sufficient precision to unambiguously determine if it is consistent with CMB observations or with quasars and radio surveys.

%%%%%%%%%%%%%%%%%%%%
\section{Discussion and conclusions}

In this work, we revisited the computation of the correlation of the $(\ell, \ell \pm 1)$ harmonic coefficients of the redshift-dependent source number counts. We found a term, which had been previously missed in the literature~\cite{Dalang2022}, and which is directly proportional to the boost factor $\beta$, without any dependence on the source population properties. This term comes from the remapping of the angular position of the sources on the sky from aberration of a moving observer. It is particularly interesting because it opens up the possibility to measure $\beta$ and the coefficient $\Fkin(z)$ independently. Using the $\fsky$ approximation and assuming Gaussianity of the galaxies distribution, we developed optimal quadratic estimators for $\beta$ and $\Fkin(z)$. We showed how measurements of the $(\ell, \ell \pm 1)$ correlations in a single redshift bin leave a significant degeneracy between $\hbeta$ and $\hF$. We then developed a multi-tracer analysis, that can be used with different populations of galaxies or different redshift bins and showed how this can be used to break the degeneracy. 

\begin{figure}[!ht]
    \begin{center}
        \includegraphics[width=.7\linewidth]{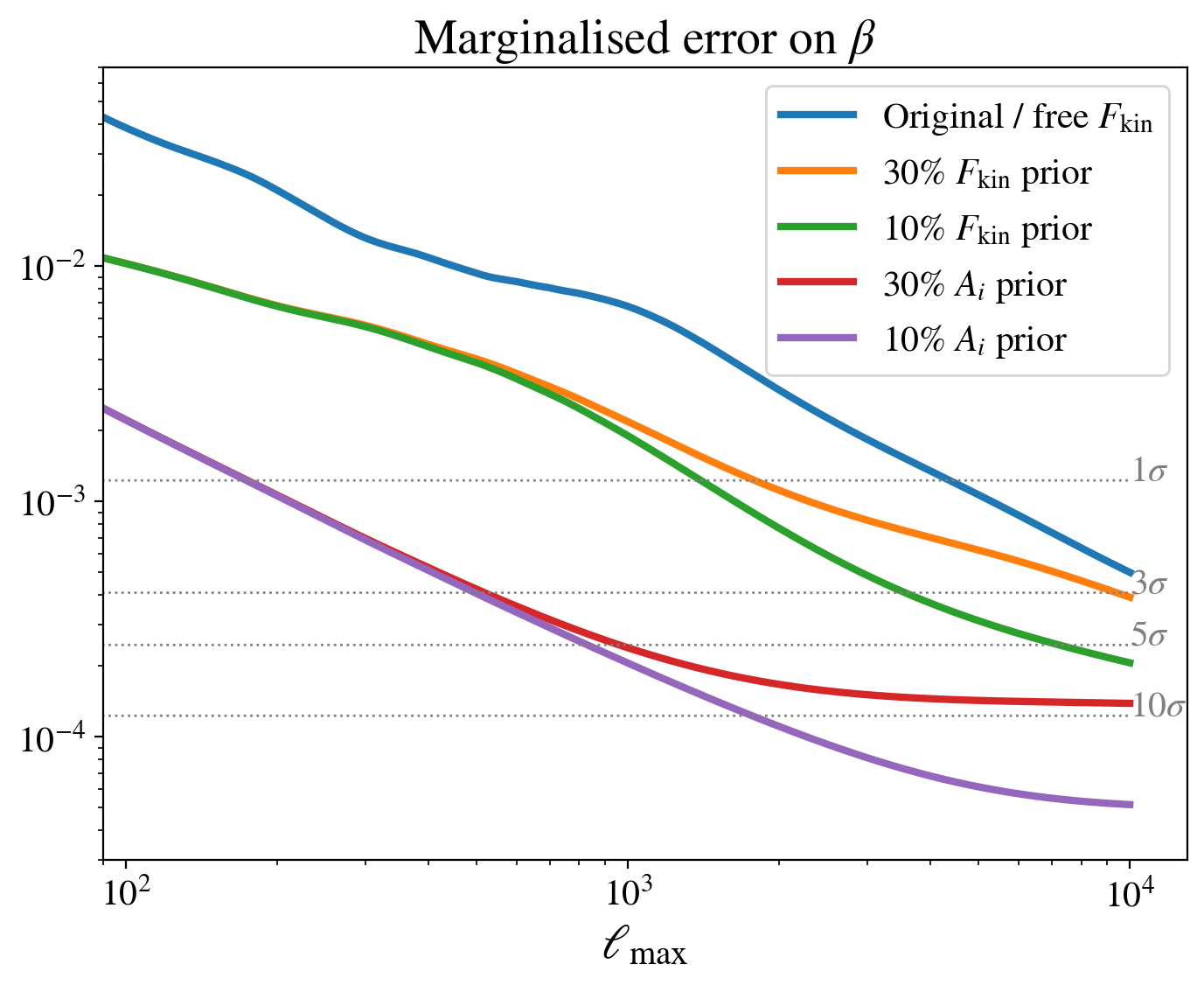}
         \caption{Summary plot: Marginalised errors on $\beta$ as a function of $\lmax$, depending on the type of multi-tracer analysis. Without any prior of the kinematic dipole, the constraint on $\beta$ is the weakest but remains detectable at nearly 3$\sigma$ at $\lmax=10,000$ with the CMB value $\beta = 1.23 \cdot 10^{-3}$. If there is $30\%$ or 10$\%$ prior knowledge of the kinematic dipole, through a direct measurement of the dipole, the constraints improve to  $3.1 \sigma$ and to $6\sigma$ at $\lmax=10,000$, respectively. In the situation where one knows to $30\%$ or $10\%$ precision the proportionality factor of Eq.\,\eqref{Eq:Dkinz}, i.e.\,$A_i = F^{(i)}_\mr{kin}/\beta$, the constraints are excellent and one can measure $\beta =1.23\cdot 10^{-3}$ with 8.9$\sigma$ or 24$\sigma$. }
        \label{Fig:summary_errors_beta}
    \end{center}
\end{figure}

Applying our technique to specifications of a \Euclid-like photometric survey, we forecasted how well $\beta$ can be measured with only one redshift bin and found that it falls short from being detectable at $1\sigma$, even when using all multipoles up to $\lmax = 10,000$. We then combined information from the 10 redshift bins of a \Euclid-like photometric sample and showed that, in this case, one can detect $\beta$ at nearly 3$\sigma$ for a fiducial value $\beta =1.23\cdot 10^{-3}$, as expected from the CMB dipole. We then showed that the detectability of $\beta$ can be strongly improved by assuming some prior knowledge on the coefficients $\Fkin^{(i)}$ in each redshift bins. In particular, assuming a 10\% prior on $\Fkin^{(i)}$ leads to a $6\sigma$ detection of $\beta$ at $\lmax=10,000$. Even more promising, prior knowledge of the evolution and magnification biases allows us to model the coefficients $\Fkin^{(i)}$ as: $\Fkin^{(i)} = A_i\,\beta$, and impose priors on the coefficients $A_i$. In this case, we showed that $\beta$ can be measured at $24\sigma$ for a 10$\%$ prior on $A_i$, which is actually a conservative assumption \cite{Maartens:2021dqy}. The results are summarized in Fig.\,\ref{Fig:summary_errors_beta}.

These potential measurements of $\beta$ with \Euclid, using $(\ell, \ell \pm 1)$ correlations in the galaxy number counts, will be paramount to better understand the tension between the CMB dipole and the dipole in radio galaxies and quasars. In particular, since the CatWISE quasar dipole measurement is roughly a factor of two larger than the expectation from the CMB dipole \cite{Secrest:2020has}, it can be interpreted as an observer's peculiar velocity measurement which is twice as large as the one from the CMB. That means that an X$\sigma$ measurement of $\beta$ can rule out one or the other interpretation at X$\sigma$, if centered on one or the other value.
This technique is thus very likely to provide the most robust and competitive constraint on the observer velocity from source number counts, confirming or not the validity of the cosmological principle.

To reach this conclusion, we made a few assumptions which are worth discussing. First, we used the popular $\fsky$ approximation to account for partial sky coverage. Techniques for a more sophisticated treatment have already been developed for the analysis of CMB lensing and can be imported to the our case. We give guidelines on what needs to be done in Appendix~\ref{App:QE-and-psky} and leave the implementation on data to future work. Since most information comes from relatively high multipoles, which are least affected by partial sky coverage, we expect only small changes. Second, the measurements of $\hbeta$ and $\hF^{(i)}$ require an input for the power spectrum $C_\ell$ up to high multipoles. In our forecasts we assumed that this was known without error. This is not the case in practice since we do not (yet) possess precise cosmological modeling of this power spectrum down to the non-linear regime. However, in Appendix~\ref{App:revising-assumptions-known-Cl} we show that this is not an issue: to the level of precision we find in this article, one can use the measured power spectrum itself (or even better any effective model, like a spline interpolation in log-log) with negligible impact on the bias and error bar of the QE. Our analysis therefore holds even when pushing to high multipoles.
Third, we used a Gaussian covariance matrix (Eq.~\eqref{Eq:Covariance_O_lm}), which might seem problematic at high multipoles in the non-linear regime. However, in Appendix~\ref{App:revising-assumptions-Gaussian-covariance}, we show that this is not an issue either: the impact of non-Gaussian covariance terms is not so much a question of multipole but instead of signal to noise. With the detection significance figures that we claim in this article we show that non-Gaussianity would have a negligible impact.

To conclude, our analysis shows that the $(\ell,\ell\pm1)$ correlations of the harmonic coefficients of the number counts, that can be measured with a \Euclid-like survey, are extremely powerful to infer the amplitude of the observer velocity $\beta$ in an unbiased and robust way. 

%%%%%%%%%%%%%%%%%%%%%%%%%%%%%%%%%%%%%%%%%%%%%%%%%%%%%%%%%%%%%%%%%%%%
\section*{Acknowledgements}
\vspace{0.2cm}

We thank Sylvain Gouyou Beauchamps for providing us with the predictions for $C_\ell^\mr{gal}$ used in the numerical computation of this article. FL and CB acknowledge support from the Boninchi foundation through the project CosmoBoost. 
CB further acknowledges financial support from the Swiss National Science Foundation and from the European Research Council (ERC) under the European Union’s Horizon 2020 research and innovation program (Grant agreement No.~863929; project title ``Testing the law of gravity with novel large-scale structure observables"). C.D.\,is supported by ERC Starting Grant SHADE (grant no.\,StG 949572).
RD acknowledges support from the Swiss National Science Foundation.
\\
The calculations in this article made use of \texttt{astropy} \citep{astropy,astropy2}, \texttt{scipy} \citep{scipy}, \texttt{numpy} \citep{numpy}, 
 and \texttt{matplotlib} \citep{Hunter2007}. 

%%%%%%%%%%%%%%%%%%%%%%%%%%%%%%%%%%%%%%%%%%%%%%%%%%%%%%%%%%%%%%%%%%%%
%%%%%%%%%%%%%%%%%%%%%%%%%%%%%%%%%%%%%%%%%%%%%%%%%%%%%%%%%%%%%%%%%%%%
%%%%%%%%%%%%%%%%%%%%          APPENDIX          %%%%%%%%%%%%%%%%%%%%
%%%%%%%%%%%%%%%%%%%%%%%%%%%%%%%%%%%%%%%%%%%%%%%%%%%%%%%%%%%%%%%%%%%%
\appendix

\section{Auxiliary calculations on spherical harmonics}
\label{app:spherical}
%%%%%%%%%%%%%%%%%%%%
\subsection{Dipole times spherical harmonic}\label{Sect:dipole-X-Ylm}
Using $Y_{10}= \sqrt{3/4\pi}\,\cos\theta$, the
following function appears in Eq.\,\eqref{Eq:alm_F} and \eqref{Eq:alm_remap}
\be
f(\bn) = \sqrt{\frac{4\pi}{3}} Y^*_{1,0}(\bn) \ Y^*_{\ell,m}(\bn)\,.
\ee
It is a function defined on the sphere, so we can decompose it in spherical harmonics. More precisely,
\ba
f(\bn) = \sum_{\ell',m'} a_{\ell',m'}^f \ Y_{\ell',m'}(\bn)\,,
\ea
with 
\ba
a_{\ell',m'}^f &= \int_{\mathbb{S}_2} \dd^2\bn \ f(\bn) \ Y^*_{\ell',m'}(\bn)
= \sqrt{\frac{4\pi}{3}} \int_{\mathbb{S}_2} \dd^2\bn Y^*_{\ell,m}(\bn) \ Y^*_{\ell',m'}(\bn) \ Y^*_{1,0}(\bn) 
= \sqrt{\frac{4\pi}{3}} G_{m,m',0}^{\ell,\ell',1}
\ea
where $G^{\ell_1,\ell_2,\ell_3}_{m_1,m_2,m_3}$ is a Gaunt coefficient. The triangular conditions imply that $a_{\ell',m'}^f$ is nonzero only if
\ba
m'=-m\,, \qquad \mr{and} \qquad \ell'=\ell \pm 1\,.
\ea
In this case, we get
\ba
a_{\ell-1,-m}^f &= \sqrt{\frac{4\pi}{3}} G_{m,-m,0}^{\ell,\ell-1,1} = \sqrt{\frac{4\pi}{3}} \sqrt{\frac{(2\ell-1)(2\ell+1)3}{4\pi}} \threejzero{\ell}{\ell-1}{1} \threej{\ell}{\ell-1}{1}{m}{-m}{0} \,,\\
a_{\ell+1,-m}^f &= \sqrt{\frac{4\pi}{3}} G_{m,-m,0}^{\ell,\ell+1,1} = \sqrt{\frac{4\pi}{3}} \sqrt{\frac{(2\ell+1)(2\ell+3)3}{4\pi}} \threejzero{\ell}{\ell+1}{1} \threej{\ell}{\ell+1}{1}{m}{-m}{0}\,.
\ea
Here, we use Wigner symbol formulas from \cite{AbramowitzStegun1972} and find that
\ba
\threejzero{\ell}{\ell-1}{1} &= (-1)^\ell \sqrt{\frac{\ell}{(2\ell-1) (2\ell+1)}} \,,\\
\threej{\ell}{\ell-1}{1}{m}{-m}{0} &= (-1)^{\ell+m} \sqrt{\frac{\ell^2-m^2}{\ell (2\ell-1)(2\ell+1)}}\,, \\
\threejzero{\ell}{\ell+1}{1} &= (-1)^{\ell+1} \sqrt{\frac{\ell+1}{(2\ell+1)(2\ell+3)}} \,,\\
\threej{\ell}{\ell+1}{1}{m}{-m}{0} &= (-1)^{\ell+1+m} \sqrt{\frac{(\ell+1)^2-m^2}{(\ell+1) (2\ell+1)(2\ell+3)}}\,,
\ea
which gives
\ba
a_{\ell-1,-m}^f &= (-1)^m \sqrt{(2\ell-1)(2\ell+1)} \sqrt{\frac{\ell}{(2\ell-1) (2\ell+1)}} \sqrt{\frac{\ell^2-m^2}{\ell (2\ell-1)(2\ell+1)}} = (-1)^m \sqrt{\frac{\ell^2-m^2}{(2\ell-1)(2\ell+1)}} \,,\\
a_{\ell+1,-m}^f &= (-1)^{m} \sqrt{(2\ell+1)(2\ell+3)} \sqrt{\frac{\ell+1}{(2\ell+1)(2\ell+3)}} \sqrt{\frac{(\ell+1)^2-m^2}{(\ell+1) (2\ell+1)(2\ell+3)}} = (-1)^{m} \sqrt{\frac{(\ell+1)^2-m^2}{(2\ell+1)(2\ell+3)}}\,.
\ea
Hence, we can write
\ba
\sqrt{\frac{4\pi}{3}} Y^*_{1,0}(\bn) \ Y^*_{\ell,m}(\bn) &= f(\bn) = (-1)^m \sqrt{\frac{\ell^2-m^2}{(2\ell-1)(2\ell+1)}} Y_{\ell-1,-m} + (-1)^{m} \sqrt{\frac{(\ell+1)^2-m^2}{(2\ell+1)(2\ell+3)}} Y_{\ell+1,-m} \nonumber \\
&= A_{\ell,m} Y^*_{\ell-1,m}(\bn) + A_{\ell+1,m} Y^*_{\ell+1,m}(\bn) \,,
\ea
where we introduced
\be
A_{\ell,m} \equiv \sqrt{\frac{\ell^2-m^2}{(2\ell-1)(2\ell+1)}}\,.
\ee
This result is used to obtain Eq.\,\eqref{Eq:alm_F_result} from Eq.\,\eqref{Eq:alm_F}.

The coefficients $A_{\ell,m}$ obey the following sum rule, which will prove useful
\ba
\sum_{m=-\ell}^{\ell} A_{\ell+1,m}^2 = \sum_{m=-\ell}^{\ell} \frac{(\ell+1)^2-m^2}{(2\ell+1)(2\ell+3)} = \frac{(\ell+1)^2(2\ell+1) - \ell(\ell+1)(2\ell+1)/3}{(2\ell+1)(2\ell+3)} = \frac{1}{3}\frac{(\ell+1)(2\ell+1)(2\ell+3)}{(2\ell+1)(2\ell+3)} = \frac{\ell+1}{3} \label{Eq:sum-A_(l+1,m)}\,.
\ea

%%%%%%%%%%%%%%%%%%%%
\subsection{Derivative of spherical harmonic}\label{Sect:Ylm-derivative}

The spherical harmonics can be defined in terms of the associated Legendre polynomicals $P_{\ell}^m$ as
\ba
Y_{\ell, m}(\theta,\phi) \equiv \sqrt{\frac{2\ell+1}{4\pi}\frac{(\ell-m)!}{(\ell+m)!}} e^{im\phi} P_\ell^m(\cos\theta)
\ea
where the normalisation is chosen for orthonormality ($\int_{\mathbb{S}_2} \dd^2\hn \ Y_{\ell, m} \ Y^*_{\ell' m'} = \delta^K_{\ell,\ell'} \ \delta^K_{m,m'}$). We want to compute the derivative $\sin\theta \ \partial_\theta  Y_{\ell, m}(\theta,\phi)$, which appears in Eq.\,\eqref{Eq:alm_remap}. To do so, we use a recurrence formula for the associated Legendre polynomials:
\ba
(1-x^2) \ \frac{\dd}{\dd x}{P_\ell^m}(x) &= \frac{1}{2\ell+1} \left[ (\ell+1)(\ell+m) \ P_{\ell-1}^m(x) - \ell(\ell+1-m) \ P_{\ell+1}^m(x) \right]
\ea
Setting $x=\cos\theta$ this gives
\ba
-\sin\theta \ \partial_\theta P_\ell^m(\cos\theta) &= \frac{1}{2\ell+1} \left[ (\ell+1)(\ell+m) \ P_{\ell-1}^m(\cos\theta) - \ell(\ell+1-m) \ P_{\ell+1}^m(\cos\theta) \right]
\ea
Thus
\ba
\sin\theta \ \partial_\theta  Y_{\ell,m} & = \frac{-1}{2\ell+1} \Bigg[ \sqrt{\frac{2\ell+1}{4\pi}\frac{(\ell-m)!}{(\ell+m)!}} (\ell+1)(\ell+m) e^{im\phi} P_{\ell-1}^m(\cos\theta) \nonumber \\
&\quad - \sqrt{\frac{2\ell+1}{4\pi}\frac{(\ell-m)!}{(\ell+m)!}} \ell(\ell+1-m) e^{im\phi} P_{\ell+1}^m(\cos\theta) \Bigg] \\
& = \frac{1}{2\ell+1} \left[ -\sqrt{\frac{2\ell+1}{2\ell-1}\frac{\ell-m}{\ell+m}} (\ell+1)(\ell+m) \ Y_{\ell-1,m} + \sqrt{\frac{2\ell+1}{2\ell+3} \frac{\ell+1+m}{\ell+1-m}} \ell(\ell+1-m) \ Y_{\ell+1,m} \right] \\
& = -\sqrt{\frac{(\ell+1)^2 \, (\ell^2-m^2)}{(2\ell-1) \, (2\ell+1)}} \ Y_{\ell-1,m} + \sqrt{\frac{\ell^2 \, \left((\ell+1)^2-m^2\right)}{(2\ell+1) \, (2\ell+3)}} \ Y_{\ell+1,m} \,,
\ea
and taking the complex conjugate, we find
\ba
\sin\theta \ \partial_\theta  Y^*_{\ell,m} &= -\sqrt{\frac{(\ell+1)^2 \, (\ell^2-m^2)}{(2\ell-1) \, (2\ell+1)}} \ Y^*_{\ell-1,m} + \sqrt{\frac{\ell^2 \, \left((\ell+1)^2-m^2\right)}{(2\ell+1) \, (2\ell+3)}} \ Y^*_{\ell+1,m} \,.
\ea
which can be simplified using $A_{\ell,m}$'s introduced in Eq.\,\eqref{Eq:A_ellm}:
\ba
\sin\theta \ \partial_\theta  Y^*_{\ell,m}(\bn) &= - (\ell+1) \, A_{\ell,m} \, Y^*_{\ell-1,m} (\bn) + \ell \, A_{\ell+1,m} \, Y^*_{\ell+1,m} (\bn)\,.
\ea
We use this expression to calculate \eqref{Eq:alm_remap}.

\section{Derivation of the Quadratic Estimators}\label{App:deriv-QE}

We recall that the off-diagonal correlators defined in Eq.\,\eqref{Eq:O_l,m} have the expectation value given in Eq.\,\eqref{eq:Exptation_Olm},
\ba\label{App:Eq:<Olm>}
\mathds{E}\left[ O_{\ell,m}(\Fkin,\beta) \right] = \Fkin \ \alpha_{\ell,m} + \beta \ \gamma_{\ell,m}\,.
\ea
The boost effect impacts linearly the (off-diagonal) correlator between two harmonic coefficients $a_{\ell,m}$. It is thus natural to attempt to build an estimator for $(\Fkin,\beta)$ that is quadratic in the $a_{\ell,m}$'s. This motivation is reinforced by the %large
striking analogy with CMB lensing, which impacts the same off-diagonal correlators (for CMB temperature and polarisation instead of galaxy counts) and for which Quadratic Estimators (QEs) have been developed with care in the literature, for instance in \cite{Hu2002}.

%%%%%
\subsection{Single parameter case}\label{App:deriv-QE-singleparam}
First, we assume that one parameter dominates in Eq.~\eqref{App:Eq:<Olm>}, e.g.\,$\Fkin \ \alpha_{\ell,m} \ll \beta \ \gamma_{\ell,m}$, so that we have a single parameter to estimate: $\beta$. This is the case where the derivation is the most similar to \cite{Hu2002}, whereas we will need to adapt calculations in later sections. We first write a general form for the quadratic estimator:
\be\label{App:Eq:QE-general-form}
\hbeta = \frac{1}{N} \sum_{\ell,m} g_{\ell,m} \ \left(O_{\ell,m}\right)^*.
\ee
The normalisation $N$ and the weights $g_{\ell,m}$ have to be optimised for the estimator to fulfill two conditions: (i) be unbiased, and (ii) have minimal variance.\\
Condition (i), i.e. $\mathds{E}\left[ \hbeta \right] = \beta$, sets the normalisation as
\be
N = \sum_{\ell,m} g_{\ell,m} \ \left(\gamma_{\ell,m}\right)^*.
\ee
Condition (ii) is fulfilled by requiring
\be
\frac{\partial \ln \mr{Var}(\hbeta)}{\partial g_{\ell,m}} = 0 \qquad \forall (\ell,m)\,,
\ee
which gives after some algebra
\be
\sum_{\ell',m'} g_{\ell',m'} \ \Sigma_{{\ell,m};{\ell',m'}} = \gamma_{\ell,m} \times \mr{cst}\,,
\ee
where $\Sigma_{{\ell,m};{\ell',m'}} \equiv \mr{Cov}\left(O_{\ell,m},O_{\ell',m'}\right)$.
After remarking that a constant multiplicative factor in $g_{\ell,m}$ cancels out with the normalisation in the estimator, we see that we can set
\be
g_{\ell,m} = \sum_{\ell',m'} \Sigma^{-1}_{{\ell,m};{\ell',m'}} \ \gamma_{\ell',m'} \,,
\ee
where $\Sigma^{-1}$ is the inverse matrix. This also sets the normalisation as
\be
N = \sum_{\ell,m;\ell',m'} \gamma_{\ell,m} \ \Sigma^{-1}_{{\ell,m};{\ell',m'}} \ \left(\gamma_{\ell',m'}\right)^*\,.
\ee
All these notations are simplified by introducing the inner product between objects with indices $(\ell,m)$
\be\label{App:Eq:dot-product}
\lbra x , y \rbra \equiv \sum_{\ell,m;\ell',m'} x_{\ell,m} \ \Sigma^{-1}_{{\ell,m};{\ell',m'}} \ \left(y_{\ell',m'}\right)^*\,,
\ee
which yields a compact equation for the estimator:
\be\label{App:Eq:QE-result-with-dot-product}
\hbeta = \frac{1}{\| \gamma\|^2} \lbra \mathbf{O} , \bm{\gamma}\rbra\,,
\ee
where $\|\bm{\gamma}\|^2 = \lbra\bm{\gamma},\bm{\gamma}\rbra$ is the norm associated with the inner product.

A posteriori, we can make the whole derivation simpler by starting with the following general form for the quadratic estimator, instead of Eq.\,\eqref{App:Eq:QE-general-form}
\be
\hbeta = \frac{1}{N} \lbra \mathbf{O} , \bm{g}\rbra\,.
\ee
In that case, the condition of unbiasedness reads $N=\lbra\bm{\gamma},\bm{g}\rbra$. Some algebra gives the variance as
\be
\mr{Var}(\hbeta) =\frac{1}{N^2} \lbra \bm{g} , \bm{g}\rbra\,,
\ee
and minimising it yields $\bm{g} \propto \bm{\gamma}$ which gives again Eq.~\eqref{App:Eq:QE-result-with-dot-product}.

%%%%%
\subsection{Two parameters case}\label{App:deriv-QE-twoparams}
In this case, we do not assume that any of the two parameters has a dominant impact on the observable, so we need to estimate them both jointly.

An easy scenario is when the templates $\bm{\alpha}$ and $\bm{\gamma}$ are orthogonal: $\lbra\bm{\alpha},\bm{\gamma}\rbra=0$. In that case, one sees that the condition of unbiasedness can be applied separately to each parameter. Thus by minimizing each variance we have two independent single-parameter QEs:
\ba
& \hF = \frac{1}{\| \bm{\alpha}\|^2} \lbra \mathbf{O} , \bm{\alpha}\rbra
& \hbeta = \frac{1}{\| \bm{\gamma}\|^2} \lbra \mathbf{O} , \bm{\gamma}\rbra \,.
\ea
Then these estimators trivially solve the following linear system
\ba
M
\times
\begin{pmatrix} \hF \\ \hbeta \end{pmatrix}
=
\begin{pmatrix} \lbra \bm{O} , \bm{\alpha} \rbra \\ \lbra \bm{O} , \bm{\gamma} \rbra \end{pmatrix}
\qquad  \mr{with} \qquad
M = \begin{pmatrix} \| \bm{\alpha} \|^2 & \lbra \bm{\alpha} , \bm{\gamma} \rbra \\ \lbra \bm{\alpha} , \bm{\gamma} \rbra & \| \bm{\gamma} \|^2 \end{pmatrix} \,.\label{App:Eq:linear-system-Dkin-beta}
\ea
The more complex scenario is the general one where $\lbra\bm{\alpha},\bm{\gamma}\rbra \neq 0$. There, the condition of unbiasedness for each parameter yields equations which are not separate anymore. Furthermore, trying to minimise the variance of each estimator separately is a bad idea because it does not account for the covariance between the two estimators, so we can end up with two estimators which are heavily correlated and thus poor marginalised error bars on $\beta$. Instead, the trick is to use transformations to reduce the computation to the independent case.\\
Eq.~\eqref{App:Eq:<Olm>} can be rewritten in vector form as $\mathds{E}\left[ \bm{O} \right] = \Fkin \, \bm{\alpha} + \beta \, \bm{\gamma}$. Thanks to the previous paragraph, we would know how to define optimal estimators if we instead had
\ba
\mathds{E}\left[ \bm{O} \right] = A_1 \, \bm{e_1} + A_2 \, \bm{e_2} \qquad \mr{with}
\qquad \lbra \bm{e_1} , \bm{e_2} \rbra = 0. \label{App:Eq:conditions-for-ei}
\ea
Indeed the estimators would be $\hat{A}_i = \frac{1}{\|\bm{e_i}\|^2} \lbra \mathbf{O} , \bm{e_i}\rbra$.
It is possible to find such vectors $(\bm{e_1},\bm{e_2})$ by orthogonalising the base $(\bm{\alpha},\bm{\gamma})$. There are (infinitely) many possible choices for this orthogonalisation, but there is one that leads to simpler calculations and has the added benefit of normalisation ($\|\bm{e_i}\|=1$): take as transfer matrix the square root of $M$. Indeed, $M$ is a real symmetric matrix so it admits a square root: a real symmetric matrix $P$ such that $M= P^2 = P P^T$.\\
Now we define $(A_1,A_2)$ and $(\bm{e_1},\bm{e_2})$ such that
\ba
\begin{pmatrix} A_1 \\ A_2 \end{pmatrix} = P \times \begin{pmatrix} \Fkin \\ \beta \end{pmatrix}
\qquad \mr{and} \qquad
\begin{pmatrix} \bm{e_1} \\ \bm{e_2} \end{pmatrix} = P^{-1} \times \begin{pmatrix} \bm{\alpha} \\ \bm{\gamma} \end{pmatrix} \,.
\ea
With a little bit of algebra, one can check that Eqs.\,\eqref{App:Eq:conditions-for-ei} are verified by this choice.\\
Following the previous paragraphs, the optimal quadratic estimators for $A_i$ are $\hat{A}_i = \frac{1}{\|\bm{e_i}\|^2} \lbra \mathbf{O} , \bm{e_i}\rbra$. We see that they solve (trivially) the linear system
\ba
\mathds{1}_{2\times2}
\times
\begin{pmatrix} \hat{A}_1 \\ \hat{A}_2 \end{pmatrix}
=
\begin{pmatrix} \lbra \bm{O} , \bm{e_1} \rbra \\ \lbra \bm{O} , \bm{e_2} \rbra \end{pmatrix}
\qquad  \mr{with} \qquad
\mathds{1}_{2\times2} = \begin{pmatrix} 1 & 0 \\ 0 & 1 \end{pmatrix}.
\label{App:Eq:linear-system-Ai}
\ea
Now we can define natural estimators for $\beta$ and $\Fkin$ as
\ba
\begin{pmatrix} \hF \\ \hbeta \end{pmatrix} = P^{-1} \times \begin{pmatrix} \hat{A}_1 \\ \hat{A}_2 \end{pmatrix}.
\ea
If we input this equation into Eq.~\eqref{App:Eq:linear-system-Ai}, some linear algebra shows that these estimators obey the same linear system as in the independent case, just with $\lbra\bm{\alpha},\bm{\gamma}\rbra \neq 0$:
\ba
\begin{pmatrix} \| \bm{\alpha} \|^2 & \lbra \bm{\alpha} , \bm{\gamma} \rbra \\ \lbra \bm{\alpha} , \bm{\gamma} \rbra & \| \bm{\gamma} \|^2 \end{pmatrix}
\times
\begin{pmatrix} \hF \\ \hbeta \end{pmatrix}
=
\begin{pmatrix} \lbra \bm{O} , \bm{\alpha} \rbra \\ \lbra \bm{O} , \bm{\gamma} \rbra \end{pmatrix}\,.
\ea
This corresponds to Eq.\,\eqref{Eq:linear-system-Fkin-beta}, the result we were looking for: a definition of the estimators that does not refer to $P$ anymore and can be entirely computed from the observable $\bm{O}$ and the galaxy power spectrum $C_\ell$ (the latter defining $\bm{\alpha}$ and $\bm{\gamma}$). 

%%%%%
\subsection{Multi parameter case}\label{App:deriv-QE-multiparams}
This section is essentially a reskin of Sec.~\ref{App:deriv-QE-twoparams} with the same derivation but with new notations that allow for an arbitrary number of $\Fkin$ parameters, as needed for the multi-tracer case.

We start with a generic equation for how the observable $O$ relates to the parameters we want to estimate
\ba
\mathds{E}\left[ O^A \right] &=  \bm{f}^A \cdot \bm{\theta} = \bm{f}^A_\mu \ \bm{\theta}^\mu,
\ea
where we used Einstein summation convention and the position of indices (up or down) is purely cosmetic. In the equation above, $\bm{\theta}$ is the vector of parameters, indexed with the Greek letter $\mu$. For example in the main text when we apply this to $n_s$ galaxy samples, we will have $\mu=0 \cdots n_s$, $\bm{\theta}_0 = \beta$ and $\bm{\theta}_j = \Fkin^{(j)}$ $\forall j=1 \cdots n_s$. The observable $O$ is indexed with a capital letter $A$. For instance with a single sample ($n_s=1$), as in Sec.~\ref{App:deriv-QE-twoparams}, $A$ would be the multi-index of spherical harmonics: $A=(\ell,m)$. In the multi-sample case, the observable can be measured for all possible pairs of samples $(j_1,j_2)$, so $A$ is a bigger multi-index: $A=(\ell,m,j_1,j_2)$. Finally, $\bm{f}^A$ dictates how the parameters impact the observable; for instance, in the case of Sec.~\ref{App:deriv-QE-twoparams}, we have
$\bm{f}^{\ell,m} = \begin{pmatrix} \alpha_{\ell,m} \\ \gamma_{\ell,m} \end{pmatrix}$.

We can now transcribe the derivation from the previous sections. First, we note $\Sigma_{A,B} \equiv \mr{Cov}\left(O^A,O^B\right)$ the covariance matrix. This allows to generalise the inner product as
\ba
\lbra \bm{x} , \bm{y}\rbra  = \left(x^A\right)^* \ \left(\Sigma^{-1}\right)_{AB} y^B\,.
\ea
We can use this inner product to define a matrix which generalises the $M$ of Sec.~\ref{App:deriv-QE-twoparams}:
\ba
M_{\mu,\nu} \equiv \lbra \bm{f}_\mu , \bm{f}_\nu\rbra = \left(f^A_\mu\right)^* \ \left(\Sigma^{-1}\right)_{AB} f^B_\nu \,.
\ea
In the following, we note simply $M \equiv \lbra \bm{f} , \bm{f}\rbra$. This is a symmetric positive matrix so it admits a square root $P$. We use $P$ to define new vectors and tensors:
\ba
\bm{A} = P \times \bm{\theta}
\qquad \mr{and} \qquad
\bm{e} = P^{-1} \times \bm{f}\,,
\ea
or more explicitely in components
\ba
\bm{A}^\mu = P^\mu_{\phantom{\mu}\nu} \ \bm{\theta}^\nu
\qquad \mr{and} \qquad
\bm{e}_\mu^A = \left(P^{-1}\right)_\mu^{\phantom{\mu}\nu} \ \bm{f}_\nu^A\,.
\ea
These tensors obey
\ba
\mathds{E}\left[ \bm{O} \right] = \bm{A} \times \bm{e}\,, \qquad \mr{i.e.} \qquad \mathds{E}\left[ \bm{O}^A \right] = \bm{A}^\mu \ \bm{e}_\mu^A\,, \qquad \mr{with} 
\qquad \lbra \bm{e_\mu} , \bm{e_\nu} \rbra = \delta^K_{\mu,\nu}\,. 
\ea
Since the $\bm{e}$ basis is orthonormal, we have optimal independent estimators for the components of the vector of parameters $\bm{A}$: $\hat{\bm{A}}^\mu = \lbra \mathbf{O} , \bm{e_\mu}\rbra$. These estimators trivially solve the linear system
\ba
\mathds{1} \times \hat{\bm{A}} = \lbra \mathbf{O} , \bm{e}\rbra. \label{App:Eq:linear-system-Amu}
\ea
When we define natural estimators for $\bm{\theta}$ as
\ba
\hat{\bm{\theta}} = P^{-1} \times \hat{\bm{A}}\,,
\ea
and we input this definition into Eq.~\eqref{App:Eq:linear-system-Amu}, some linear algebra shows that these estimators obey
\ba
\lbra \bm{f}, \bm{f} \rbra \cdot \hat{\bm{\theta}} = \lbra \bm{O}, \bm{f} \rbra\,,
\ea 
which corresponds to Eq.\,\eqref{Eq:syseq_multi}, the result that we were looking for: a definition of optimal estimators solely in terms of the starting quantities $\bm{O}$ and $\bm{f}$. 

%%%%%%%%%%%%%%%%%%%%
\section{Quadratic estimator and partial sky coverage}\label{App:QE-and-psky}

In this Appendix we show how to generalise the quadratic estimators derived in the main text in order to account for the survey partial sky coverage. To do so, we follow the formalism developed for Quadratic Estimators of CMB lensing, which translates straightforwardly to our case with minor modifications. Specifically we use methods and notations similar to \cite{Planck2015-lensing}. The process is composed of two steps: (i) Wiener filtering, and (ii) construction of the QE and mean field subtraction. The steps are explained below, with a summary at the end.\\

\noindent \textit{(i) Wiener filtering}\\
We call $d$ the data, i.e. the masked map. It is considered as a vector of length set by the number of pixels. We call $s$ the signal, i.e.\,the full sky galaxy map, and $S$ its covariance (which is diagonal in harmonic space). Following \cite{Planck2015-lensing}, we then write
\be
d = s + n\,,
\ee
where $n$ is a fictious independent noise which has infinite variance ($N^{-1}=0$) in masked pixels. The goal of this fictious noise is to downweight to zero all masked pixels in the Wiener filtering. $N$ is diagonal in pixel space: $N_{p,p'} = \delta^K_{p,p'} N_p$ where $p$ is the pixel index.
What value should we give to $N_p$ in observed (non-masked) pixels? In principle it should be set to zero, but in practice it is useful to set it to a small value $\epsilon>0$ so that we can invert and get $N^{-1}$ in all pixels.
Then our Wiener-filtered data $d^{WF}$ (Ref.\,\cite{Planck2015-lensing} calls it $\overline{s}$) is the inverse covariance weighting of $d$:
\be
d^{WF} = W d\,,  \quad \mr{where} \quad W = \left[ S+N \right]^{-1} = S^{-1} \left[ S^{-1} + N^{-1} \right]^{-1} N^{-1}\,,
\ee
where the rewriting on the right hand side allows to only need $N^{-1}$ and not $N$, as the latter formally has infinite values. The first multiplication $N^{-1} d$ downweights masked pixels to zero, which is what we want.\\

\noindent\textit{(ii) Construction of the QE and mean field subtraction}\\
We should now build our estimator only in terms of $d^{WF}$.
So our observable is
\be
O_{\ell,m} = d^{WF}_{\ell,m} \ (d^{WF}_{\ell+1,m})^*\,.
\ee
Its expectation value is
\ba
\mathds{E}\left[ O_{\ell,m} \rbra  &= \lbra (W d)_{\ell,m} (W d)_{\ell+1,m}^* \right] \\
&= \mathds{E}\left[ (W s)_{\ell,m} (W s)_{\ell+1,m}^* \right] + \mathds{E}\left[ (W n)_{\ell,m} (W n)_{\ell+1,m}^* \right] \\
&= \mathds{E}\left[ (W s^\mr{boost})_{\ell,m} (W s^\mr{boost})_{\ell+1,m}^* \right] + \mathds{E}\left[ (W s^\mr{unboost})_{\ell,m} (W s^\mr{unboost})_{\ell+1,m}^* \right] + \mathds{E}\left[ (W n)_{\ell,m} (W n)_{\ell+1,m}^* \right] \\
& \approx F_{\ell} \, F_{\ell+1} \ \mathds{E}\left[ O_{\ell,m}^\mr{fullsky} \right] + \mathds{E}\left[ O_{\ell,m}^\mr{unboost} \right]_\mr{MC}\,,
\ea
where in the second line, we assumed that the signal and noise are 
uncorrelated. In the third line, we separated the galaxy clustering contributions coming from the boost (of order $\mathcal{O}(\beta^1)$) and the unboosted classical part (of order $\mathcal{O}(\beta^0)$) that now leaks in the estimator due to isotropy breaking by the mask. Finally, in the fourth line, we (i) used the "diagonal" approximation from \cite{Planck2015-lensing}, $(W s)_{\ell,m} \simeq F_\ell \ s_{\ell,m}$, i.e.\,we neglect the small amount of mask-induced mode-mixing, and (ii) took the unboosted contributions (from both the signal and noise) through Monte-Carlo (MC) average over (unboosted) simulations. This latter MC average is called the "mean-field" in the lensing literature. Following that tradition, we call
\be
O_{\ell,m}^{MF} \equiv \mathds{E}\left[ O_{\ell,m}^\mr{unboost} \right]_\mr{MC}\,,
\ee
and our new observable is the mean-field subtracted $O'_{\ell,m} \equiv O_{\ell,m} - O_{\ell,m}^{MF}$. Then we have
\ba
\mathds{E}\left[ O'_{\ell,m} \right] &= F_{\ell} \, F_{\ell+1} \ A_{\ell+1,m} \times \Big[ \Fkin \left( C_\ell + C_{\ell+1} \right) - \beta \ \left( (\ell+2) \, C_{\ell+1} - \ell \, C_\ell \right) \Big] \\
&= \phantom{f}^T \bm{f}_{\ell,m} \cdot \bm{\theta}\,,
\ea
with the vector of parameters $\bm{\theta} = (\beta, \Fkin)$. Hence, we are in a position similar to the fullsky case and we need only a final element to build the estimator: the covariance of $O'$:
\ba
\Cov\left(O'_{\ell,m},O'_{\ell',m'}\right) & = \mathds{E}\left[ (W d)_{\ell,m} (W d)_{\ell+1,m}^* \ (W d)_{\ell',m'} (W d)_{\ell'+1,m's}^* \right]_c \\
& \approx F_{\ell}^2 \, F_{\ell+1}^2 \ C_{\ell} \, C_{\ell+1} \ \delta^K_{\ell,\ell'} \, \delta^K_{m_,m'} + \mathcal{O}(\beta)\,,
\ea
where we used the diagonal approximation again.\\
Then we can use the same developments as in Sec.~\ref{Sect:estimators-single-tracer} and \ref{Sect:estimators-multi-tracer}: we can define an inner product
\ba
\lbra \mathbf{x} , \mathbf{y} \rbra \equiv \sum_{\ell,m} \frac{\left(x_{\ell,m}\right)^* \ y_{\ell,m} }{F_{\ell}^2 \, F_{\ell+1}^2 \ C_{\ell} \, C_{\ell+1}} \,.
\ea
and the optimal quadratic estimators are given by 
\ba
\bm{\hat{\theta}} = \lbra \bm{f}, \bm{f} \rbra^{-1} \times \lbra \bm{O}', \bm{f} \rbra .
\ea

\noindent \textit{(iii) Summary}\\
In summary, compared to the full sky case the differences are
\begin{itemize}
    \item the raw observable is measured on Wiener-filtered maps
    \item the mean-field has to be subtracted from the observable
    \item the average and covariance of $\bm{O}'$ are modified with Wiener filter factors leading to different inner products. For instance $\lbra \bm{O}', \bm{f} \rbra$ writes explicitely as
    \ba
    \lbra \bm{O}', \bm{f} \rbra = \sum_{\ell,m} \frac{\left(O'_{\ell,m}\right)^* \ F_{\ell} \, F_{\ell+1} \; A_{\ell+1,m}}{F_{\ell}^2 \, F_{\ell+1}^2 \ C_{\ell} \, C_{\ell+1}}
    \begin{pmatrix} - (\ell+2) \, C_{\ell+1} + \ell \, C_\ell \\ C_\ell + C_{\ell+1} \end{pmatrix}
    \ea    
\end{itemize}

%%%%%%%%%%%%%%%%%%%%
\section{Revising assumptions: Gaussian covariance and known power spectrum}\label{App:revising-assumptions}

In this section, we examine the validity of two assumptions that we made in the main text 
and that might seem questionable due to non-linearities on small scales: (i) using a Gaussian covariance, and (ii) that we know perfectly the angular power spectrum $\Cl$ needed as an ingredient for the covariance and the templates $\bm{\alpha}$ and $\bm{\gamma}$. We do so by quantifying the impact of going beyond these assumptions, and we will show that this impact can be neglected to the degree of precision achievable by our method.

%%%%%%%%%%
\subsection{Gaussian covariance}\label{App:revising-assumptions-Gaussian-covariance}

When we push the analyses to high $\lmax$, the non-Gaussian contribution (coming from the trispectrum) to the covariance of a quantity quadratic in the galaxy density becomes inevitably more important. In fact this non-Gaussianity has a non-negligible impact on cosmological constraints using the galaxy power spectrum already for $\lmax=300$, as shown in \cite{Lacasa2016}. This non-Gaussian contribution is complex, being composed of a large number of terms \cite{Lacasa2018b}, although Super-Sample Covariance (SSC) is generally the term with the dominant impact \cite{Lacasa2020}.

However, in our case we are not interested in cosmological constraints using measurements of the $\Cl$'s. Instead we are interested in obtaining a detection of $\beta$ with the $(\ell,\ell+1)$ correlator. That makes a large difference, because the impact of non-Gaussianity is not so much dictated by $\lmax$ but much more by the statistical power of the measurement. Indeed, \cite{Lacasa2019} showed that the impact of SSC on parameter constraints scale with the (square of the) signal to noise ratio (S/N) of the measurement. More precisely, in the low S/N regime the \emph{relative} difference of the Fisher matrix element is \cite{Lacasa2019}
\ba
\left| \frac{\delta F}{F} \right| = \cos^2\theta \times \frac{(S/N)^2}{(S/N)_\mr{max}^2}
\ea
where $\cos\theta$ is a quantity defined in \cite{Lacasa2019} that quantifies how much the inferred parameter is degenerate with the impact of SSC. As hinted by the notation this number lies in the range $-1 \leq \cos\theta \leq 1$. Here, we are going to be conservative and take the upper limit $\cos^2\theta=1$. Then $(S/N)_\mr{max}$ is the SSC-induced plateau of signal to noise. For the median \Euclid{} redshift bin, Ref.\,\cite{Lacasa2019} found a value of $(S/N)_\mr{max} \simeq 250$. We are going to be extremely conservative and use this value for our best possible scenario: the multi-tracer analysis of all 10 \Euclid{} bins. This is extremely conservative because the impact of SSC scales down with the survey volume. With these assumptions, for the best possible scenario where we claim a detectability of $24\sigma$, the dominant non-Gaussian impact would increase our error bars by \emph{at most}
\ba
\left| \frac{\delta F}{F} \right| \leq \left(\frac{24}{250}\right)^2 \sim 1\%.
\ea
We thus conclude that the impact of non-Gaussianity can safely be neglected at the level of precision we claim in the main text.

%%%%%%%%%%
\subsection{Known power spectrum \texorpdfstring{$\Cl$}{Cl}}\label{App:revising-assumptions-known-Cl}

The coefficients $\alpha_{\ell, m}$ and $\gamma_{\ell, m}$ that enter in our estimators depend on the power spectra $C_\ell$ (see Eqs.\,\eqref{eq:alpha_lm}-\eqref{eq:gamma_lm}). As discussed in the main text, one can either use measured power spectra or theoretically modelled power spectra. Here, we show that the optimal solution is a "hybrid model", where one uses the theoretically modelled spectra at low $\ell$, in the perturbative regime, and the measured spectra at high $\ell$, when the measurement uncertainties are negligible, but no reliable model exists. More precisely, we use the fact that
\begin{itemize}
\item at the highest multipoles, $\lmax\sim 10,000$ we enter the shot-noise dominated regime. In this regime, the power spectrum is known: $\Cl = 1/N_\mr{gal}$ where $N_\mr{gal}$ is the actual number of galaxies (per steradian) present in the catalog. So this number is known exactly (see e.g. the shot-noise subtraction discussion in \cite{Lacasa2018b}). Hence, we have perfect knowledge of $\Cl$ there,
\item at low multipoles a perturbative model can be used for a faithful reproduction of the Baryonic Acoustic Oscillations (BAO). Such models are already available in the literature due to the cosmological importance of the BAO. They have sub-percent accuracy, which we will show below to be largely sufficient for our purpose,
\item at intermediate multipoles which lie in the non-linear regime but are not yet dominated by shot-noise, for our purpose we do not need a cosmological model of $\Cl$. Instead, the raw (binned) measurement of $\Cl$ can be used. Indeed it can easily reach percent level accuracy, which we will show below to be largely sufficient for our purpose.
\end{itemize}

To justify our second and third points, we now quantify the issues raised by an incorrect $\Cl$ and show each time that percent level precision is enough for our purpose. To do so, we first justify the second point by considering the impact of a fixed incorrect $\Cl$, due to the (small) imprecision of the galaxy clustering model. Next we justify the third point by considering a stochastic $\Cl$ measured from the galaxy clustering data itself. In both cases, we will show the calculations in the simpler case of a single tracer and single parameter estimate of $\beta$.\\

First is the case of a fixed incorrect $\Cl$. Here the QE has two potential issues: bias and increase of error bar. Concerning the bias, one sees that an incorrect covariance matrix does not bias the QE ; instead the bias comes exclusively from using an incorrect template $\bm{\gamma}_\mr{mod}$ different from the true template $\bm{\gamma}_\mr{true}$. Some algebra shows that the expectation value of the QE is
\ba
\mathds{E}\left[\hbeta\right] = \frac{\lbra \bm{\gamma}_\mr{true}, \bm{\gamma}_\mr{mod}\rbra}{\lbra \bm{\gamma}_\mr{mod}, \bm{\gamma}_\mr{mod}\rbra} \times \beta
\ea
Hence an x\% precision on $\Cl$ (and thus on $\bm{\gamma}$) yields a bias on $\beta$ of at most x\%. In the best scenario in the main text we had 4-5\% precision on $\beta$, which means that we need a precision on $\Cl$ of a few \%. The latter is well within reach of current models, see e.g. \cite{DonaldMcCann2022}.\\
Concerning the increase of error bar, one sees that it comes mostly from using an incorrect covariance matrix $\Sigma_\mr{mod}$ different from the true covariance $\Sigma_\mr{true}$. Some algebra shows that the variance of the QE is
\ba
\mr{Var}\left(\hbeta\right) = \frac{\bm{\gamma}^T \, \Sigma_\mr{mod}^{-1} \,  \Sigma_\mr{true}^{\phantom{-1}} \, \Sigma_\mr{mod}^{-1} \, \bm{\gamma}}{\|\bm{\gamma}\|^4}.
\ea
This does not simplify analytically in the general case, except if we take the relative variation of the power spectrum to be small and constant with multipole, in which case we get
\ba
\frac{\delta\mr{Var}\left(\hbeta\right)}{\mr{Var}\left(\hbeta\right)} = 2 \frac{\delta \Cl}{\Cl}.
\ea
This means that a precision on $\Cl$ of a few \% also gives a precision on the error bar on $\beta$ of a few percent, which is  sufficient for our needs.\\
In conclusion to this first case, we see that it is a viable option for our purpose to use current non-linear galaxy modeling in order to produce a template for $\Cl$, in the range of multipoles where the precision of the model is below a few percent.\\

Second, we examine the case where we use the data itself to estimate $\Cl$. We consider this a superior alternative, in terms of both precision and practicality, in the high multipole regime outside of the BAO. This case is in principle more complex because we have to account for the fact that stochastic variations in the template $\bm{\gamma}$ and the estimated covariance $\Sigma$ are correlated together as well as with fluctuations of the observable $O_{\ell,m}$ that we are fitting against. This affects the two parts of the estimator: the normalisation $\|\bm{\gamma}\|^2$ and the inner product $\lbra \bm{O}, \bm{\gamma}\rbra$. But we can argue that the relative variation of both these quantities is of the same order as the relative variation of the power spectrum and then determine conditions for the latter to be small enough.\\
We begin with the normalisation ; with an estimated power spectrum $\hat{C}_\ell$, it becomes
\ba
\|\bm{\gamma}\|^2_\mr{est} &= \sum_\ell \frac{\ell+1}{3} \frac{\left( (\ell+2) \, \hat{C}_{\ell+1} - \ell \, \hat{C}_\ell \right)^2}{\hat{C}_{\ell} \,\hat{C}_{\ell+1}} \\
&\simeq \sum_\ell \frac{\ell+1}{3} \frac{\left( (\ell+2) \, {C}_{\ell+1} - \ell \, {C}_\ell \right)^2}{{C}_{\ell} \, {C}_{\ell+1}} \times \left[ 1 - \frac{\delta \Cl}{\Cl} - \frac{\delta C_{\ell+1}}{C_{\ell+1}} +2 \frac{\delta \left((\ell+2) \, {C}_{\ell+1} - \ell \, {C}_\ell\right)}{(\ell+2) \, {C}_{\ell+1} - \ell \, {C}_\ell} \right] \,,
\ea
where the variations $\delta X$ are now stochastic variables. Noting that for a local power law $\Cl \propto \ell^{-n}$ we have
\ba
(\ell+2) \, {C}_{\ell+1} - \ell \, {C}_\ell \simeq \frac{1}{\ell} \frac{\dd \ell^2 \Cl}{\dd \ell} = (2-n) \Cl\,,
\ea
and thus that its relative variation is the same as $\frac{\delta \Cl}{\Cl}$, we can put the upper limit on the standard deviation
\ba
\frac{\sigma(\|\bm{\gamma}\|^2)}{\|\bm{\gamma}\|^2} \leq 4 \left|\frac{\sigma(\Cl)}{\Cl}\right|\,.
\ea
Next we have the inner product ; with an estimated power spectrum $\hat{C}_\ell$ it becomes
\ba
\lbra \bm{O}, \bm{\gamma}\rbra_\mr{est} &= \sum_{\ell,m} - O_{\ell+1,m} \ A_{\ell+1,m} \ \frac{ (\ell+2) \, \hat{C}_{\ell+1} - \ell \, \hat{C}_\ell }{\hat{C}_{\ell} \,\hat{C}_{\ell+1}} \fsky \\
&\simeq \sum_{\ell,m} - O_{\ell+1,m} \ A_{\ell+1,m} \ \frac{ (\ell+2) \, C_{\ell+1} - \ell \, C_\ell }{C_{\ell} \,C_{\ell+1}} \fsky \times \left[ 1 - \frac{\delta \Cl}{\Cl} - \frac{\delta C_{\ell+1}}{C_{\ell+1}} + \frac{\delta \left((\ell+2) \, {C}_{\ell+1} - \ell \, {C}_\ell\right)}{(\ell+2) \, {C}_{\ell+1} - \ell \, {C}_\ell} \right] \,,
\ea
with the same reasoning as for $\|\bm{\gamma}\|^2$, we get the upper limit
\ba
\frac{\sigma(\lbra \bm{O}, \bm{\gamma})\rbra}{\lbra \bm{O}, \bm{\gamma}\rbra} \leq 3 \left|\frac{\sigma(\Cl)}{\Cl}\right|
\ea
Thus, for both parts of the QE, the increase of error bar is under control for our precision requirements if the standard deviation of the estimated power spectrum is less than a percent: $\sigma(\Cl)/\Cl \leq 1\%$.\\
The calculation above stands for the increase of error bar. Remains the potential issue of the bias. Calculating the bias requires to develop the normalisation and inner product to second order, since $\mathds{E}(\delta \Cl)=0$ but $\mathds{E}\left((\delta \Cl)^2\right)\neq 0$. This is cumbersome and not particularly enlightening ; instead it suffices to remark that the resulting relative biases will necessarily be proportional to $\left(\sigma(\Cl)/\Cl\right)^2$. Thus, if we apply the condition $\sigma(\Cl)/\Cl \leq 1\%$ to keep control of the error bars, we will \emph{a fortiori} keep control of the bias.\\
To close the argument we thus need to find the requirements for $\sigma(\Cl)/\Cl \leq 1\%$ and show that they are realistic. For the input $\Cl$ we simply use the actual binned power spectrum measurement.\footnote{We note that this could easily be refined. Indeed, in this regime the spectra are expected to be very smooth, close to a power law with a slowly varying spectral index. This could be used to refine a more precise estimate, e.g. with spline interpolation in the space $(\ln\ell,\ln\Cl)$.} There, the standard $\fsky$ formula for the cosmic variance gives
\ba
\left(\frac{\sigma(\Cl)}{\Cl}\right)^2 = \frac{2}{(2\ell+1) \ \Delta\ell \ \fsky}\,.
\ea
For coming galaxy surveys like \Euclid{} we have $\fsky\sim\sfrac{1}{3}$. Thus for a bin width $\Delta\ell=100$, our requirement is fulfilled at a multipole as low as $\ell=300$. Since the power spectrum is expected to be modeled accurately down to at least $\ell=2000$ \cite{Euclid-IST}, we see that we can easily patch the two regimes: cosmological model at low multipoles and actual measurement at high multipoles.\\
We conclude that the main text assumption of a known power spectrum is largely valid for our precision requirements.

%%%%%%%%%%
\subsection{Non-linear growth and bias}\label{App:revising-assumptions-NLbg}

We start by recalling the equation for the boosted counts, Eq.~\eqref{Eq:Nfinal}, isolating the $\Fkin$ term:
\ba
N(z,\bn,S_*) =& \left [1+\cos\theta \, D_\mr{kin}(z)\right ] \tilde{\bar N}(z, S_* ) + \left [1+\beta\sin\theta\,\partial_\theta\right ]\tilde{\bar N}(z, S_* ) \ \delta_\mr{g}(z,\bn,S_*)\,  +\cos\theta \, \Fkin(\bn,z) \, \tilde{\bar N}(z, S_* ) \ \delta_\mr{g}(z,\bn,S_*),
\ea
where
\ba
\Fkin(\bn,z) &= D_\mr{kin}(z)+\left[\frac{\partial \ln\delta_\mr{g}}{\partial\ln (1+z)}(\bn,z)+ \frac{2}{r\HH}\frac{\partial \ln\delta_\mr{g}}{\partial\ln S_*}(\bn,z)\right] \beta
\ea
In the main text, the log derivative reduced to the growth rate or bias derivatives which were assumed scale independent. This assumption cannot be sustained in practice in the non-linear regime. To study this, we rewrite the count equation as
\ba
N(z,\bn,S_*) =& \left [1+\cos\theta \, D_\mr{kin}(z)\right ] \tilde{\bar N}(z, S_* ) + \left [1+\beta\sin\theta\,\partial_\theta\right ]\tilde{\bar N}(z, S_* ) \ \delta_\mr{g}(z,\bn,S_*)\,  + \cos\theta \, \tilde{\bar N}(z, S_* ) \ \mathcal{D}_k \;\! \delta_\mr{g}(z,\bn,S_*),
\ea
where we defined the differential operator
\ba
\mathcal{D}_k = D_\mr{kin} + \beta \left[\frac{\partial }{\partial\ln (1+z)}+ \frac{2}{r\HH}\frac{\partial }{\partial\ln S_*}\right] = D_\mr{kin} + \beta \mathcal{D}.
\ea
This can be propagated into the derivation of Sec.~\ref{Sect:boost-observables}, where it only changes Eq.~\eqref{Eq:alm_F_result} into
\ba
\delta a_{\ell,m}^\mr{F} &= \tilde{\bar{N}}(z) \int_{\mathbb{S}_2} \dd^2 \bn \, \mathcal{D}_k \;\! \dg(z,\bn)  \left[ A_{\ell,m} \, Y^*_{\ell-1,m}(\bn) + A_{\ell+1,m} \, Y^*_{\ell+1,m}(\bn) \right] \\
&= A_{\ell,m} \ a_{\ell-1,m}\left(\tilde{\bar{N}} \ \mathcal{D}_k \;\! \dg\right) + A_{\ell+1,m} \ a _{\ell+1,m}\left(\tilde{\bar{N}} \ \mathcal{D}_k \;\! \dg\right) .
\ea
In Sec.~\ref{Sect:stat-impact}, the effect of non-linear growth and bias impacts only Eq.~\eqref{Eq:Olmaverage}, that it modifies into
\ba
\mathds{E}\left[ O_{\ell,m} \right] 
&= A_{\ell+1,m} \cdot \left[ \Big(C_\ell^{\dg,\mathcal{D}_k \;\!\dg} + C_{\ell+1}^{\dg,\mathcal{D}_k \;\!\dg}\Big) - \beta \ \Big( (\ell+2) \, C_{\ell+1}^{\dg,\dg} - \ell \, C_\ell^{\dg,\dg} \Big)\right] \, ,
\ea 
while previously only $C_\ell=C_\ell^{\dg,\dg}$ appeared.\\
If we define $\Fkin(\ell)$ as the ratio of the two different power spectra
\ba
\Fkin(\ell) \equiv \frac{C_\ell^{\dg,\mathcal{D}_k \;\!\dg}}{C_\ell^{\dg,\dg}} = D_\mr{kin} + \beta \frac{C_\ell^{\dg,\mathcal{D} \;\!\dg}}{C_\ell^{\dg,\dg}}\,,
\ea
we see that we get and equation similar to Eq.~\eqref{Eq:Olmaverage}
\ba
\mathds{E}\left[ O_{\ell,m} \right] 
&= A_{\ell+1,m} \cdot \left[ \Big(\Fkin(\ell) \ C_\ell + \Fkin(\ell+1) \ C_{\ell+1}\Big) - \beta \ \Big( (\ell+2) \, C_{\ell+1} - \ell \, C_\ell \Big)\right]\,.
\ea 
In practice, the ratio $\left. C_\ell^{\dg,\mathcal{D} \;\!\dg} \middle/ C_\ell^{\dg,\dg} \right.$ should be modeled using a non-linear description of the large scale structure, if a sufficiently precise one is available, or using simulations.\footnote{A small number of simulations should be sufficient because the ratio will be prone to cosmic variance cancellation. Indeed upwards (downwards) fluctuations of matter density will increase (decrease) both the numerator and the denominator of the ratio.} Part of the ratio can also be determined from the survey itself, namely the derivative $\frac{\partial}{\partial \ln S_*}$ which may be the hardest to model or simulate accurately, by taking several flux cuts and estimating the derivative with finite difference. From this, we would have a parametric model for $\Fkin(\ell)$, whose parameters are constrained by priors from the model/simulations/survey.\\
In order to study the impact of introducing such a parametric model for $\Fkin(\ell)$, we choose here to test as toy model a simple agnostic form:
\ba
\Fkin(\ell) = \Fkin^{[0]} + \Fkin^{[1]} \times \frac{\ell}{\ell_0}\,, \label{App:Eq:Fkin(ell)-model}
\ea
with $\ell_0$ an arbitrary pivot scale. In our implementation, we chose $\ell_0=\ell_\mr{max}=10,000$ so that both parameters have the same order of magnitude: $\left|\Fkin^{[1]}\right|=\left|\Fkin^{[0]}\right|$ corresponds to a 100\% variation of $\Fkin$ over the multipole range.
For our test to be conservative, we decide to assume no prior on any of the two parameters, i.e. both $\Fkin^{[0]}$ and $\Fkin^{[1]}$ are free.\\
Some straightforward generalisation of the derivation in Appendix~\ref{App:deriv-QE} shows that we can build joint Quadratic Estimators for $\beta$ and the $\Fkin$ parameters. For instance, in the single tracer case, we have 
\ba
\begin{pmatrix}
\| \bm{\alpha} \|^2 & \lbra \bm{\alpha} , \bm{\gamma} \rbra & \lbra \bm{\alpha} , \bm{\delta} \rbra \\
\lbra \bm{\gamma}, \bm{\alpha}  \rbra & \| \bm{\gamma} \|^2 & \lbra \bm{\gamma} , \bm{\delta} \rbra \\
\lbra \bm{\delta} , \bm{\alpha} \rbra & \lbra \bm{\delta} , \bm{\gamma} \rbra & \| \bm{\delta} \|^2
\end{pmatrix}
\times
\begin{pmatrix} \hF^{[0]} \\ \hbeta \\ \hF^{[1]} \end{pmatrix}
=
\begin{pmatrix} \lbra \bm{O} , \bm{\alpha} \rbra \\ \lbra \bm{O} , \bm{\gamma} \rbra \\ \lbra \bm{O} , \bm{\delta} \rbra \end{pmatrix} \,,
\ea
where the weights are
\ba
\alpha_{\ell,m} = A_{\ell+1,m} \, \left(C_\ell + C_{\ell+1}\right) \qquad
\gamma_{\ell,m} = -A_{\ell+1,m} \, \left( (\ell+2) \, C_{\ell+1} - \ell \, C_\ell \right) \qquad
\delta_{\ell,m} = A_{\ell+1,m} \, \left(\ell \, C_\ell + (\ell+1) \, C_{\ell+1}\right)  .
\ea
The multi-tracer case is obtained through a straightforward generalization.\\
We implemented these estimators and reran the analysis of Sec.~\ref{Sect:forecast} with the new free parameter $\Fkin^{[1]}$. We found that it yields a modest increase of the error bars on $\beta$, with the impact on the detection significance at $\lmax=10,000$ summarised in Table~\ref{App:Tab:impact-of-Fkin^[1]}.

\begin{table}[!th]
\begin{center}
\begin{tabular}{c|c|c|c}
case & constant $\Fkin$ & scale-dependent $\Fkin$ & relative difference  \\
\hline
Original / free $\Fkin$  & 2.5 $\sigma$ & 2.0 $\sigma$ & 19\% \\
30\% $\Fkin$ prior  & 3.1 $\sigma$ & 2.6 $\sigma$ & 18\% \\
10\% $\Fkin$ prior & 6.0 $\sigma$ & 5.0 $\sigma$ & 16\% \\
30\% $A_i$ prior & 8.9 $\sigma$ & 8.7 $\sigma$ & 2\% \\
10\% $A_i$ prior & 24 $\sigma$ & 21 $\sigma$ & 11\% \\
\end{tabular}
\end{center}
\caption{\textit{First column:} all cases studied in the main text, following the nomenclature in the legend of Fig.~\ref{Fig:summary_errors_beta}. \textit{Second column:} detection significance for $\beta$ at $\lmax=10,000$ when $\Fkin$ is constant with multipole, i.e. the results quoted in the main text. \textit{Third column:} detection significance for $\beta$ at $\lmax=10,000$ when $\Fkin$ is allowed to vary with $\ell$ through Eq.~\eqref{App:Eq:Fkin(ell)-model}, and we marginalise over that scale dependence. \textit{Fourth column:} relative difference between the second and third columns.}
\label{App:Tab:impact-of-Fkin^[1]}
\end{table}

We see that the impact of this conservative non-linear toy model is small, ranging between 2\% and 20\%. We found the impact to generally be even smaller when the multipole range is cut to a smaller $\lmax$. We conclude that the impact of non-linear growth and bias will not alter significantly the conclusions in the main text, and in particular the prospect of a significant detection of $\beta$ with the proposed method.

%%%%%%%%%%%%%%%%%%%%%%%%%%%%%%%%%%%%%%%%%%%%%%%%%%%%%%%%%%%%%%%%%%%%
%%%%%%%%%%%%%%%%%%%%        BIBLIOGRAPHY        %%%%%%%%%%%%%%%%%%%%
%%%%%%%%%%%%%%%%%%%%%%%%%%%%%%%%%%%%%%%%%%%%%%%%%%%%%%%%%%%%%%%%%%%%

\bibliographystyle{JHEP}
\bibliography{bibliography}

%%%%%%%%%%%%%%%%%%%%%%%%%%%%%%%%%%%%%%%%%%%%%%%%%%%%%%%%%%%%%%%%%%%%
%%%%%%%%%%%%%%%%%%%%            END             %%%%%%%%%%%%%%%%%%%%
%%%%%%%%%%%%%%%%%%%%%%%%%%%%%%%%%%%%%%%%%%%%%%%%%%%%%%%%%%%%%%%%%%%%

\end{document}